\documentclass[twocolumn,showpacs,preprintnumbers,amsmath,amssymb]{revtex4}
\usepackage{graphics,epsfig,subfigure}
\usepackage{color}
\usepackage[linktocpage]{hyperref}

\begin{document}
\renewcommand{\baselinestretch}{1.3}
\newcommand\beq{\begin{equation}}
\newcommand\eeq{\end{equation}}
\newcommand\beqn{\begin{eqnarray}}
\newcommand\eeqn{\end{eqnarray}}
\newcommand\nn{\nonumber}
\newcommand\fc{\frac}
\newcommand\lt{\left}
\newcommand\rt{\right}
\newcommand\pt{\partial}
\newcommand\tx{\text}
\newcommand\mc{\mathcal}

\allowdisplaybreaks

\title{Thick branes in Born-Infeld determinantal gravity in Weitzenb\"ock spacetime}
\author{
Ke Yang$^a$\footnote{keyang@swu.edu.cn},
Hao Yu$^{b}$\footnote{yuhaocd@cqu.edu.cn},
Yi Zhong$^{c}$\footnote{zhongy@hnu.edu.cn, corresponding author}
}

\affiliation{
$^{a}$School of Physical Science and Technology, Southwest University, Chongqing 400715, China\\
$^{b}$Physics Department, Chongqing University, Chongqing 401331, China\\
$^{c}$School of Physics and Electronics Science,\\
Hunan Provincial Key Laboratory of High-Energy Scale Physics and Applications, Hunan University, Changsha 410082, P. R. China}

\begin{abstract}
By adopting the idea of Born-Infeld electromagnetism, the Born-Infeld determinantal gravity in Weitzenb\"ock spacetime provides a way to smooth the Big Bang singularity at the classical level. We consider a  thick braneworld scenario in the higher-dimensional extension of this gravity, and investigate the torsion effects on the brane structure and gravitational perturbation. For three particular parameter choices, analytic domain wall solutions are obtained. They have a similar brane configuration that the brane thickness becomes thinner as the spacetime torsion gets stronger.  For each model, the massless graviton is localized on the brane with the width of localization decreasing with the enhancement of the spacetime torsion, while the massive gravitons propagate in the bulk and contribute a correction term proportional to ${1}/{(k r)^{3}}$ to the Newtonian potential. A sparsity constraint on the fundamental 5-dimensional gravitational scale is estimated from the gravitational experiment. Moreover, the parameter ranges in which the Kaluza-Klein gravitons are tachyonic free are analyzed.
\end{abstract}

\maketitle

\section{Introduction}

Einstein's general relativity (GR) is the cornerstone of modern cosmology, which provides the most accurate descriptions of a variety of phenomena in our universe. In GR, the gravitation is described by the curvature of spacetime, where the affine connection is symmetric and coincides uniquely with the Levi-Civita connection. A well-known alternative gravity theory dynamically equivalent to GR is the so-called teleparallel gravity \cite{Cho1976,Hayashi1979}, which allows us to interpret GR as a gauge theory for a translation group.
In teleparallel gravity, the spacetime is characterized by the curvature-free Weitzenb\"ock
connection rather than the torsion-free Levi-Civita connection, and the dynamical field is the vielbein instead of metric in Weitzenb\"ock spacetime \cite{Maluf2013,Bahamonde2021}. Following the spirit of the popular $f(R)$ gravity which generalizes the Lagrangian of GR to be some functions of the Ricci scalar $R$, the $f(T)$ gravity generalizes the Lagrangian of  teleparallel gravity to be some functions of the torsion scalar $T$ \cite{Ferraro2007,Bengochea2009,Linder2010}. This impactful modification of teleparallel gravity provides possible explanations for the accelerating expansion of the universe without invoking the dark energy \cite{Bengochea2009,Cardone2012} and for the inflation without resorting to the inflaton field \cite{Ferraro2007,Ferraro2008}.

It is well-known that GR suffers from the inevitable singularities at the beginning of Big Bang and the center of black holes \cite{Hawking1970}. Some attempts to solve the singularity problem of GR at the classical level were done by replacing the Einstein-Hilbert action with the Born-Infeld-type determinantal action \cite{Deser1998,Vollick2004,Banados2010,Chen2016}. The Born-Infeld determinantal gravity (BIDG) in Weitzenb\"ock spacetime was also considered in Refs.~\cite{Ferraro2010,Fiorini2013,Fiorini2016}. This theory leads to second-order field equations and supports some regular cosmological solutions by replacing the possible initial singularity with a de-Sitter phase or a bounce. The tensor evolution of these cosmological solutions can hold stability in a large parameter space in the early universe, and the theoretical parameter $\lambda$ is constrained by the speed of gravitational waves \cite{Yang2019a}. However, other cosmological singularities such as Big Rip, Big Bang, Big Freeze, and Sudden singularities may emerge in some regions of parameter space \cite{Bouhmadi-Lopez2014a}.  The Schwarzschild geometry was considered in this framework as well \cite{Fiorini2016a}.

The idea that our spacetime may have hidden extra dimensions in ultraviolet regime is a long historical topic since the proposal of Kaluza-Klein (KK) theory in the 1920s.  As the most plausible candidate for unifying all fundamental interactions, the superstring/M-theory requires extra dimensions for mathematical consistency. Instead of traditional compact extra dimension scenario, the braneworld scenario suggests that our universe is a 3-brane embedded into a higher-dimensional bulk, which opens a new way to solve some long-standing problems in particle physics and cosmology, such as the gauge hierarchy problem and cosmological constant problem \cite{Arkani-Hamed1998,Arkani-Hamed2000,Randall1999,Randall1999a,Arvanitaki2016}. Thick brane scenario is considered as a smooth generalization of Randall-Sundrum-2 (RS2) model \cite{Randall1999a}. The most interesting thick brane configuration is the domain wall, which is a global topological defect in a multidimensional bulk \cite{Dzhunushaliev2010}. The domain wall brane could be a potential carrier, onto which the Standard Model could possibly be transplanted \cite{Davies2008}.

The projected Gauss-Codazzi equations of RS2 model in the 5-dimensional teleparallel gravity contain two extra terms arising from the extra degrees of freedom in the teleparallel Lagrangian \cite{Nozari2013}. Furthermore, the brane cosmology provides an equivalent viewpoint as RS2 model in GR, however, the projected effect on the brane is determined by the projection of torsion tensor \cite{Geng2014a}. The inflation and dark energy dominated stage were realized on the brane by generalizing RS2 model in 5-dimensional $f(T)$ gravity \cite{Bamba2013}. Analytic thick brane solutions were obtained in the 5-dimensional $f(T)$ gravity \cite{Yang2012b,Menezes2014,Yang2018,Wang2018a}. The gravitational perturbation against the thick branes and the corresponding resonance spectrum were investigated \cite{Guo2016,Tan2021}. Thick string-like brane-world models were constructed in the 6-dimensional $f(T)$ gravity, where the torsion effects on the models were analyzed as well \cite{Moreira2021}.

In this work, we will consider the thick brane scenario in the higher-dimensional BIDG in Weitzenb\"ock spacetime and analyze the torsion effects on the brane structure and gravitational perturbation. The layout of the paper is as follows: In Sec.~\ref{BIDG}, the higher-dimensional BIDG in Weitzenb\"ock spacetime is briefly introduced. In Sec.~\ref{Brane_Scenario}, the thick brane scenario is constructed in $(d+1)$-dimensional BIDG, and the first-order formalism is employed in order to solve the field equations analytically. Specifically, for $d=4$, analytical 3-brane solutions are obtained for three particular parameter choices in Sec.~\ref{Solutions}. In Sec.~\ref{Perturbation}, the gravitational perturbation against the domain wall backgrounds is discussed. Finally, brief conclusions and discussions are presented. Throughout the paper, the capital Latin indices $A, B,C \cdots$ and small Latin indices $a, b,c \cdots$ label the $(d+1)$-dimensional and  $d$-dimensional coordinates of tangent space, respectively, while the capital Latin letters $K,L,M, N, \cdots$ and Greek letters $\mu,\nu, \rho,\sigma\cdots$ label the $(d+1)$-dimensional and $d$-dimensional spacetime indices, respectively.

\section{Higher-dimensional BIDG in Weitzenb\"ock spacetime}\label{BIDG}

In Weitzenb\"ock spacetime, the fundamental dynamic field is the  vielbein ${e^{A}}_{M}$, which refers to the metric through the relation $g_{MN}={e^{A}}_{M}{e^{B}}_{N}\eta_{AB}$, with $\eta_{AB} = \text{diag}(-1, 1, \cdots,1)$ the Minkowski metric for the tangent space. The torsion tensor ${T^{P}}_{MN}={\Gamma^{P}}_{NM}-{\Gamma^{P}}_{MN}$ is constructed in terms of the
Weitzenb\"ock connection ${\Gamma^{P}}_{MN}={e_{A}}^{P}\pt_{N}{e^{A}}_{M}$. With the torsion tensor, the contorsion tensor is defined as ${K^{P}}_{MN}=\fc{1}{2}({{T_{M}}^{P}}_{N}+{{T_{N}}^{P}}_{M}-{{T^{P}}_{M}}_{N})$. By defining the superpotential torsion tensor ${S_P}^{MN}=\fc{1}{2}(K^{MN}{}_{P}+\delta^{N}_{P}T_{Q}{}^{QM}-\delta^{M}_{P}T_{Q}{}^{QN})$, the torsion scalar is given by $T={S_P}^{MN}{T^{P}}_{MN}$.

The action of $(d+1)$-dimensional BIDG in Weitzenb\"ock spacetime takes the form \cite{Fiorini2013}
\beqn
S=-\frac{\lambda}{16\pi G_{d+1}}\int{}d^{d+1}x\left[\sqrt{-|g_{MN}+{2}{\lambda}^{-1} F_{MN}|}
\right.\nonumber \\
\left.
-\sqrt{-|g_{MN}|} \right]
+\int{}d^{d+1}x \sqrt{-|g_{MN}|} \mc{L}_\tx{M},
\label{Full_Action}
\eeqn
where the rank-2 tensor $F_{MN}$ is a function of vielbein field ${e^{A}}_{M}$ and its derivatives, $\lambda$  the Born-Infeld constant with mass dimension 2, and $\mc{L}_\tx{M}$ the Lagrangian of matter fields. The $(d+1)$-dimensional gravitational constant $G_{d+1}$ will be set to $4\pi G_{d+1}=1$ for later convenience.

In the low-energy limit  $\lambda\rightarrow\infty$, the above action approximates to
\beq
S\approx-\frac{1}{4}\int{}d^{d+1}x ~ e \tx{Tr}(F_{MN})+\int{}d^{d+1}x ~e \mc{L}_\tx{M},
\eeq
where $e=|{e^A}_M|=\sqrt{-|g_{MN}|}$. Thus, the teleparallel gravity can be recovered in the case of $\text{Tr}(F_{MN})=T$. Generally, $F_{MN}$ is given by $F_{MN}=\alpha F^{(1)}_{MN}+\beta F^{(2)}_{MN}+\gamma F^{(3)}_{MN}$ with $F^{(1)}_{MN}={S_M}^{PQ}T_{NPQ}$, $F^{(2)}_{MN}={S_{PM}}^{Q}{T^{P}}_{NQ}$, and $F^{(3)}_{MN}=g_{MN}T$ \cite{Fiorini2013}. The three dimensionless parameters $\alpha$, $\beta$ and $\gamma$ satisfy the condition $\alpha+\beta+(d+1)\gamma=1$ in order to yield $\text{Tr}(F_{MN})=T$.

By varying the action with respect to the vielbein, one gets the Euler-Lagrange equation
\beq
\frac{\pt \mathcal{L}_\tx{G}}{\pt e^A{}_M}-\pt_S\lt(\frac{\pt \mathcal{L}_\tx{G}}{\pt (\pt_S e^A{}_M)}\rt)=\frac{4 e}{\lambda }{\Theta_A}^M,
\eeq
with each term written explicitly as
\beqn
\frac{\pt \mathcal{L}_\tx{G}}{\pt e^A{}_M}&=&\frac{|\mc{U}_{KL}|^{\frac{1}{2}}(\mc{U}^{-1})^{QP} }{2} \lt(e_{A(P} \delta^M{}_{Q)}+\frac{2 \pt F_{PQ}}{\lambda  \pt e^A{}_M}\rt)
\nonumber \\
&-&{e_A}^M e,\\
\frac{\partial \mathcal{L}_\tx{G}}{\pt (\pt_S e^A{}_M)}&=&\frac{|\mc{U}_{KL}|^{\frac{1}{2}}(\mc{U}^{-1})^{QP}}{\lambda}\frac{ \pt F_{PQ} }{\pt (\pt_S e^A{}_M)},\\
{\Theta_A}^M &=& \fc{1}{e}\fc{\pt  (e\mc{L}_\tx{M})}{\pt e^A{}_M},
\eeqn
where $\mathcal{L}_\tx{G}$ represents the gravitational Lagrangian and  $\mc{U}_{KL}=g_{KL}+2\lambda^{-1} F_{KL}$.
After contracting the index $A$ of tangent space via multiplying a vielbein ${e^A}_N$, the equations of motion read \cite{Fiorini2016a}
\beqn
&&\frac{|\mc{U}_{KL}|^{\frac{1}{2}} \lt({\mc{U}}^{-1}\rt)^{QP}  }{2}\lt[{\delta^M}_{(P} g_{NQ)} + \frac{2e^A{}_{N}}{\lambda} \frac{\partial F_{PQ}}{ \partial {e^A}_M} \rt]-{\delta^M}_N e
\nn\\
&&-\frac{e^A{}_{N}}{\lambda }\partial_S\lt[ |\mc{U}_{KL}|^{\frac{1}{2}}  \lt({\mc{U}}^{-1}\rt)^{QP} \frac{\partial F_{PQ}}{\partial(\partial_S {e^A}_M)}\rt]\!=\!\frac{4 e}{\lambda }{\Theta_N}^M,
\label{EoM}
\eeqn
where the symbol $(~)$ denotes the symmetric tensor components. The energy-momentum tensor ${\Theta}_{N}{}^{M}={e^A}_N {\Theta_A}^M$ is symmetric and conserved if the action of the matter fields is local Lorentz invariant \cite{Li2011d}. With some algebra, these two partial derivative terms are written explicitly as
\beqn
\frac{\partial F_{PQ}}{ \partial {e^A}_M}&=&\alpha
\lt({\delta^M}_P F^{(1)}_{AQ}+{\delta^{M}}_{Q} F^{(1)}_{P A}
\right.\nn\\
&+&\left.
{Q^M}_{APKL} {T_Q}^{KL}-2S_{PK(A}{T_Q}^{K M)}\rt)\nn\\
&+&\beta \lt({Q^M}_{AKPL} {T^K}_{Q}{}^L - {S_{KP}}^{(M}{T^K}_{Q A)} \rt)
\nn\\
&+&\gamma \lt({\delta^M}_{(P} e_{AQ)}T-4g_{PQ}{F^{(2)M}}_{A}\rt),\\
\frac{\partial F_{PQ}}{\partial(\partial_S {e^A}_M)}&=&\alpha\lt(2e_{AQ} {S_P}^{SM}+ D_{PKL A}{}^{[SM]} {T_Q}^{KL} \rt)
\nn\\
&+&
\beta\lt({S_{AP}}^{[M} {\delta ^{S]}}_Q + D_{K PL A}{}^{[SM]} {T^{K}}_Q{}^L \rt)\nn\\
&+&4 \gamma g_{PQ} {S_{A}}^{SM},
\eeqn
where $[~]$ denotes the skew-symmetric tensor components, and the tensors ${D^C}_{PQ B}{}^{KL}$ and ${Q^K}_A{}^C{}_{PQ}$ are expressed as
\begin{align}
{D^C}_{PQ B}{}^{KL}&=\fc{1}{4}\lt({\delta_P}^K {\delta_Q}^L {\delta_B}^C-e^{CL}e_{B[P}\delta_{Q]}{}^K \rt)
\nn\\
&+\fc{1}{2}{e_B}^{L}{e^C}_{[P}\delta_{Q]}^K,\\
{Q^K}_A{}^C{}_{PQ}&=\frac{1}{4}\lt(e^{CK}T_{[PQ]A}-{\delta^K}_{[P}{T_{AQ]}}^C\rt)
\nn\\
&
-\fc{1}{2}\lt({\delta_A}^C {\delta^K} _{[P}
{T^L}_{L Q]}-{e^C}_{[P}{T^K}_{AQ]}\rt).
\end{align}

\section{Thick brane scenario and first-order formalism}\label{Brane_Scenario}

To investigate the braneworld scenario, the $(d+1)$-dimensional line element is described by
\beq
ds^2=a^{2}(y)\eta_{\mu\nu}dx^\mu dx^\nu+dy^2,
\eeq
where $y$ denotes the extra dimension coordinate perpendicular to the brane, and $a(y)$ is the so-called warp factor. Correspondingly, the proper vielbein reads ${e^A}_M=\tx{diag}\lt(a(y),\cdots,\right.$ 
$\left.a(y),1 \rt)$.

In order to construct a braneworld configuration, a single background scalar field is included, of which the Lagrangian reads
\beq
\mathcal{L}_{\tx{M}}=-\frac{1}{2}\pt^M\phi\pt_M\phi-V(\phi).
\eeq
The scalar field should depend on the coordinate $y$ only, to be consistent with the $d$-dimensional Poincar\'e invariance of the   metric ansatz. Then, the energy-momentum tensor is written explicitly as
\beqn
Q^\mu{}_\nu &=&\left(-\frac{1}{2}\phi '^2 -V\right)\delta ^{\mu }{}_{\nu },\nn\\
Q^y{}_y &=&\frac{1}{2}\phi '^2-V,
\eeqn
where the prime denotes the derivative with respect to the coordinate $y$. Since the brane is constructed by the scalar field, the brane configuration is more easily seen from its effective energy density, which is defined as \cite{Liu2012,Yang2012a}
\beq
\rho(y)=-Q^0{}_0-V_0,
\eeq
 where $V_0$ is the vacuum energy of the scalar potential.

With the vielbein and matter energy-momentum tensor, the field equations (\ref{EoM}) become
\begin{subequations}\label{EoM_Exp}
\beqn
&&\frac{\left(1-B H^2\right)^\fc{d-2}{2}}{(1-A H^2)^\fc{1}{2}}\left[1+(d-1)B H^2-dA B H^4\right]-1
\nn\\
&&=\frac{4 }{\lambda }\left(\frac{\phi '^2}{2}-V\right),\qquad\label{EoM_Exp1}\\
&&\frac{\left(1-B H^2\right)^\fc{d-4}{2}}{\left(1-A H^2\right)^\fc{3}{2}}\bigg[1+\frac{A+d B}{d}H'-(A-(d-2)B)H^2
\nn\\
&&-\frac{2(2 d+1) A + d(d-1) B}{d}BH^2 H'-(d + 1) A^2 B^2 H^6 H' \nn\\
&&-(2 (d - 1) A+ (d - 1) B)B  H^4+(d A+ (2 d - 1) B)AB  H^6
\nn\\
&&+\fc{(2 d^2 + 1) B + 3 d A}{d} ABH^4 H' - d A^2 B^2 H^8\bigg]-1
\nn\\
&&=\frac{4 }{\lambda }\left(-\frac{\phi '^2}{2}-V\right),\quad\label{EoM_Exp2}\\
&&\phi''+4 H \phi'  = V_\phi,\qquad\qquad\qquad~\label{EoM_Exp3}
\eeqn
\end{subequations}
where $H\equiv a'/a$, $V_\phi\equiv \frac{dV}{d\phi}$, $A=d (d-1) (\beta +2 \gamma )/\lambda$, and $B=(d-1) (2 \alpha +\beta +2 d \gamma)/\lambda$.

It is noted that the system is underdetermined since there are only two of the three equations of motion are independent, but we have three unknown variables $a$, $\phi$, and $V(\phi)$. In order to solve the system analytically, we employ the first-order formalism \cite{DeWolfe2000,Gremm2000a,Afonso2006}, which can transform the equations of motion (\ref{EoM_Exp}) into first-order equations by introducing a superpotential $W(\phi)$, namely,
\begin{subequations}\label{EoM_Superpotential}
\beqn
H\!&\!=\!&\!-\fc{W}{3},\label{EoM_H}\\
\phi'\!&\!=\!&\!\fc{\lambda }{4 d}\fc{(9-B W^2)^{\frac{d-4}{2}}}{3^d(9-A W^2)^\fc{3}{2}}\Big[
729 (A+dB)-81  ((4 d+2)A
\nn\\
&+& (d-1) d B)B W^2+9 (3 A d+2 B d^2+B)A B W^4
\nn\\
&-& d (d+1)A^2 B^2 W^6\Big] W_\phi,\label{EoM_dphi}\\
V\!&\!=\!&\!\fc{\lambda }{4}-\frac{\lambda }{3^{d+1}}\bigg[\frac{\left(9-B W^2\right)^{\frac{d-2}{2}}}{4 (9-A W^2)^\fc{1}{2}} \big[81+9 (d-1)B W^2\nn\\
&-&  d A B W^4\big]+\frac{(9-B W^2)^{\fc{d-4}{2}}}{8 d (9-A W^2)^\fc{3}{2}}\Big[2187 (A+d B)\nn\\
&-&243\big[(4 d+2)A+ d(d-1) B\big] B W^2\nn\\
&+&27 \big[3 d A + (2 d^2+1)B \big] A B W^4
\nn\\
&-&3 d (d+1) A^2 B^2  W^6\Big]\phi'  W_\phi \bigg].\label{EoM_V}
\eeqn
\end{subequations}
where $W_\phi\equiv \fc{dW}{d\phi}$.

In order to recover the massless chiral fermions on the brane, the bulk should be $Z_2$-symmetric along the coordinate $y$. Thus, the warp factor $a(y)$ must be an even function of extra dimension. From Eq.~(\ref{EoM_H}), the superpotential $W(\phi)$ must be chosen as an odd function of the scalar $\phi$. With an appropriate ansatz of superpotential, the variables can be solved directly from the above equations.

\section{Analytic 3-brane solutions}\label{Solutions}

Specifically, we focus on solving the 3-brane solutions in a 5-dimensional bulk ($d=4$) in the rest of the work. The higher-dimensional brane solutions can be solved similarly. However, Eqs.~(\ref{EoM_H}) and (\ref{EoM_dphi}) can not be integrated out generally due to their complicated form. Therefore, it is convenient to consider some particular cases by fixing the values of free parameters, which can simplify the equations greatly.

\subsection{Case $A=B$}

The first interesting case is $A=B$. Then from $A=12(\beta +2 \gamma )/\lambda$, $B=3 (2 \alpha +\beta +8\gamma)/\lambda$, and $\alpha+\beta+5\gamma=1$, the parameters are fixed as $\alpha = \frac{3}{5}-3 \gamma$ and $\beta = \frac{2}{5}-2 \gamma$, yet $\gamma$ left as a free parameter. It leads to $A=B=\frac{24}{5 \lambda }$. So the Eqs.~(\ref{EoM_Superpotential}) reduce to
\begin{subequations}\label{EoM_Superpotential_AB}
\beqn
a' \!&\!=\!&\! -\fc{aW}{3},\\
\phi' \!&\!=\!&\! \lt(\frac{1}{2}-\frac{16 W^2}{15 \lambda } \rt)\lt(1-\frac{8 W^2}{15 \lambda }\rt)^\fc{1}{2} W_\phi,\\
V \!&\!=\!&\! \fc{\lambda }{4}-\lt(\frac{\lambda }{4}+\frac{2 W^2}{5}-\frac{64 W^4}{225 \lambda } -\frac{ \phi ' W_\phi}{4}+\frac{8  \phi ' W_\phi W^2}{15 \lambda }\rt)
\nn\\
\!&\! \times \!&\! \lt(1-\frac{8 W^2}{15 \lambda }\rt)^\fc{1}{2} .
\eeqn
 \end{subequations}

\begin{figure*}[htb]
\begin{center}
\subfigure[$~a(y)$]  {\label{Fig_Solution_AB_Warp_Factor}
\includegraphics[width=7cm,height=5cm]{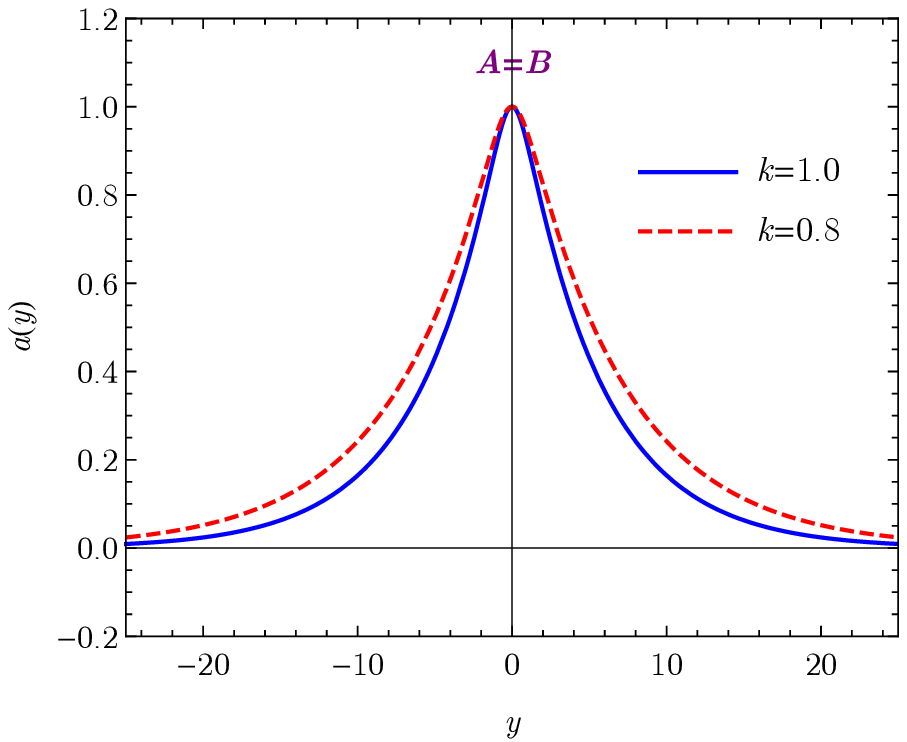}}
\subfigure[$~\phi(y)$]  {\label{Fig_Solution_AB_Scalar_Field}
\includegraphics[width=7cm,height=5cm]{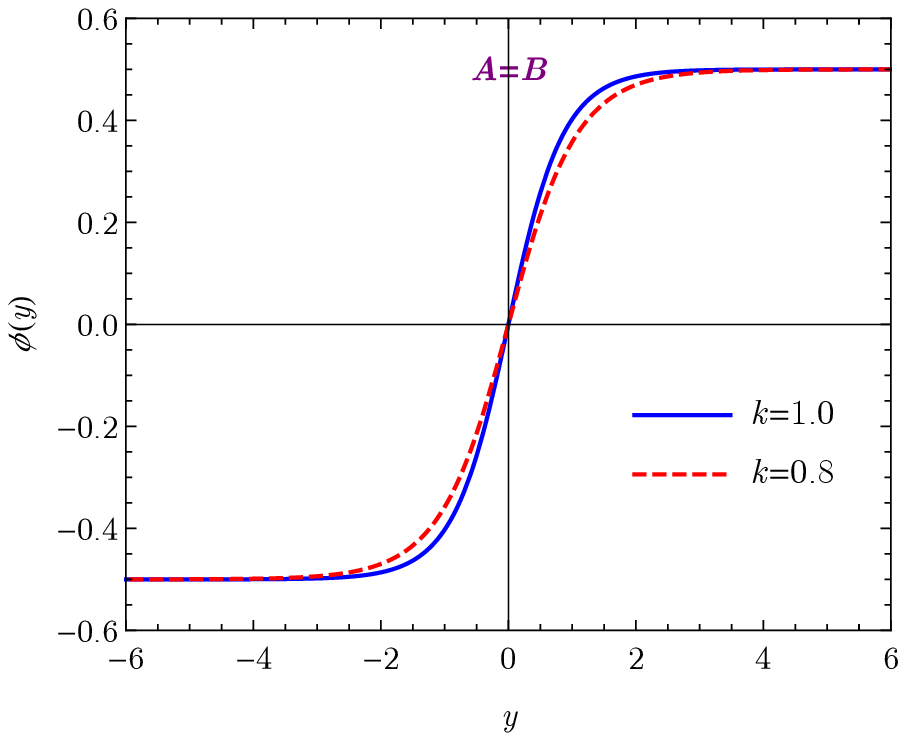}}
\subfigure[$~V(\phi)$]  {\label{Fig_Solution_AB_Scalar_Potential}
\includegraphics[width=7cm,height=5cm]{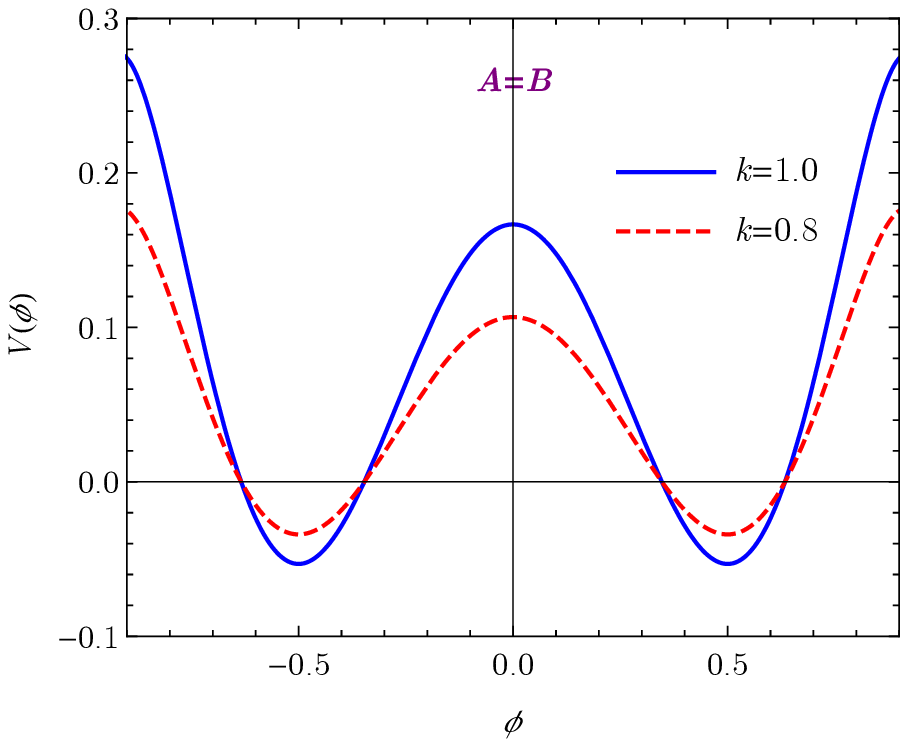}}
\subfigure[$~\rho(y)$]  {\label{Fig_Solution_AB_Energy_Density}
\includegraphics[width=7cm,height=5cm]{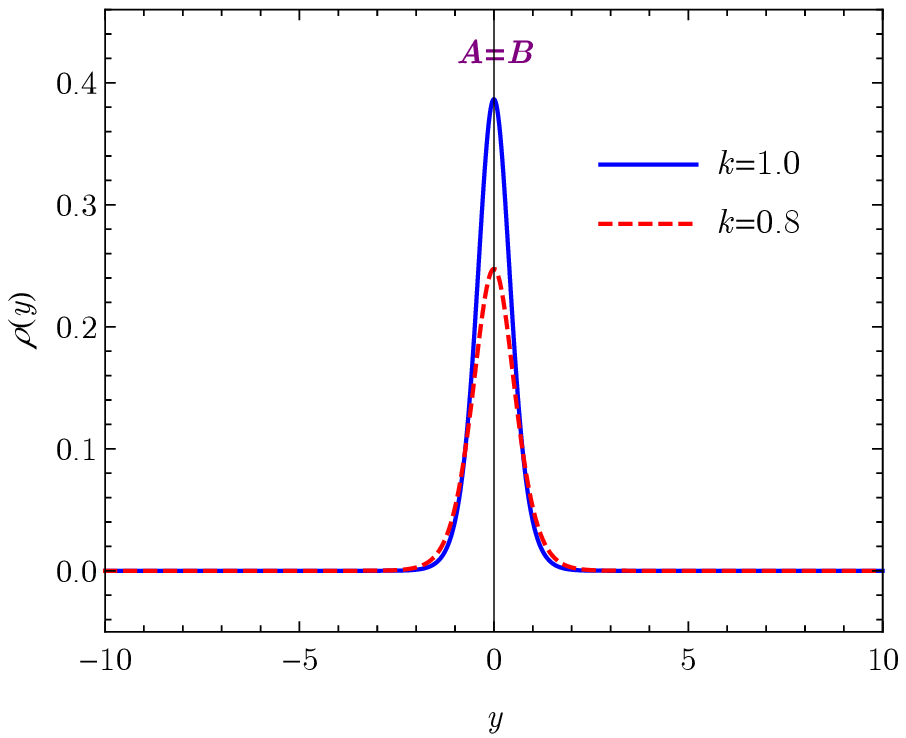}}
\end{center}
\caption{The profiles of the warp factor $a(y)$, scalar field $\phi(y)$, scalar potential $V(\phi)$, and energy density $\rho(y)$ for the case of $A=B$.}
\label{Fig_Solution_AB}
\end{figure*}

A set of analytical solutions can be obtained with the superpotential ansatz $W(\phi)=\fc{2k}{\sqrt{3}}\phi$, where the mass dimension one parameter $k$ is defined as $\lambda\equiv\frac{32}{45}k^2$.  Then taking $W(\phi)$ into Eqs.~(\ref{EoM_Superpotential_AB}), the set of analytical solutions are obtained as
 \begin{subequations}\label{Solution_AB}
 \beqn
a(y) \!&\!=\!&\!\fc{\text{sech} ^{\frac{1}{3 \sqrt{3}}}(k y) }{\lt[\big(2-\sqrt{3}\big) \lt( 2+\sqrt{4-\text{sech}^2(k y)}\rt)\rt]^{\frac{1}{3 \sqrt{3}}}},\label{Solution_AB_a}\\
 \phi (y) \!&\!=\!&\!\frac{\tanh (k y)}{\sqrt{3+\tanh ^2(k y)}},\\
V( \phi) \!&\!=\!&\! \frac{k^2}{90}  \Big[\phi ^2\left(4 \phi ^2-3\right) \lt(45-60 \phi ^2+16 \sqrt{1-\phi ^2}\rt) \nn\\
&-&16 \sqrt{1-\phi ^2}+31\Big],
 \eeqn
\end{subequations}
where the integration constants have been chosen such that $a(0)=1$ and $\phi(0)=0$. The solutions of the warp factor $a(y)$, scalar field $\phi(y)$, and scalar potential $V(\phi)$ are illustrated  in Figs.~\ref{Fig_Solution_AB_Warp_Factor}, \ref{Fig_Solution_AB_Scalar_Field}, and \ref{Fig_Solution_AB_Scalar_Potential}. As shown in  Fig.~\ref{Fig_Solution_AB_Warp_Factor}, the warp factor holds the $Z_2$ symmetry and has a typical bell shape profile. As $y\to \pm \infty$, the scalar field $\phi (\pm \infty)\to\pm\fc{1}{2}$, which just correspond to the two local minima of the scalar potential, i.e., $V_0=V(\pm\fc{1}{2})=(8-6 \sqrt{3}) k^2/45$. Since the scalar field non-trivially maps the boundaries of extra dimension into two scalar vacua, it is a kink soliton \cite{Dzhunushaliev2010}. 

 With the solution (\ref{Solution_AB_a}), the torsion scalar reads $T=-\fc{16 k^2 \tanh^2(k y)}{9 \lt(4-\text{sech}^2(k y)\rt)}$. It approaches $-\fc{4 k^2}{9}$ at the vacua $y\to\pm\infty$.  Since the absolute value of the torsion scalar is monotonically increasing with the parameter $k$ (or $\lambda$), the torsion of spacetime is enhanced as the parameter $k$ goes larger. As shown in Fig.~\ref{Fig_Solution_AB_Energy_Density}, the energy density of the brane is localized at the origin of the extra dimension, and the brane thickness becomes thinner as the background torsion gets stronger.

By choosing the free parameter $\gamma=1/5$, one has $\alpha =\beta = 0$. Then, the action (\ref{Full_Action}) reduces to a Born-Infeld-$f(T)$-type one \cite{Fiorini2013,Yang2018}, i.e.,
\beqn
S_{\text{BI}}&=&-\frac{1}{4}\int{}d^5x ~ ef(T)\nn\\
&=&-\frac{\lambda}{4}\int{}d^5x ~ e \lt[\lt(1+\frac{2T}{5\lambda}\rt) ^{5/2}-1\rt].
\eeqn
In addition, the same analytic solutions (\ref{Solution_AB}) were obtained and the issue about trapping fermions on the domain wall was considered in Ref.~\cite{Yang2018}.

 \subsection{Case $A=0$}
\begin{figure*}[htb]
\begin{center}
\subfigure[$~a(y)$]  {\label{Fig_Solution_A0_Warp_Factor}
\includegraphics[width=7cm,height=5cm]{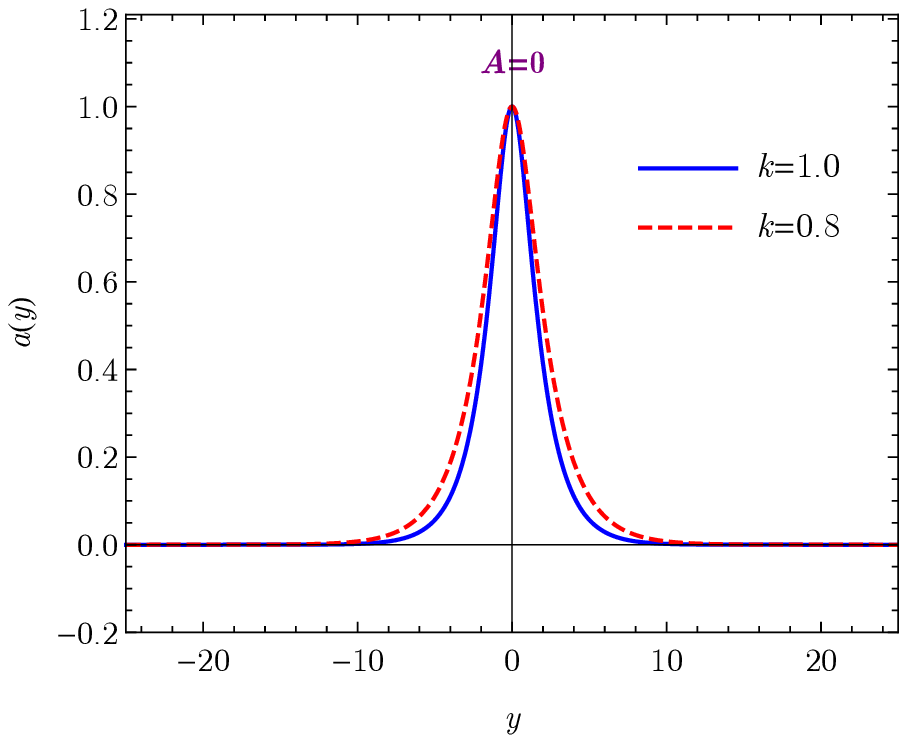}}
\subfigure[$~\phi(y)$]  {\label{Fig_Solution_A0_Scalar_Field}
\includegraphics[width=7cm,height=5cm]{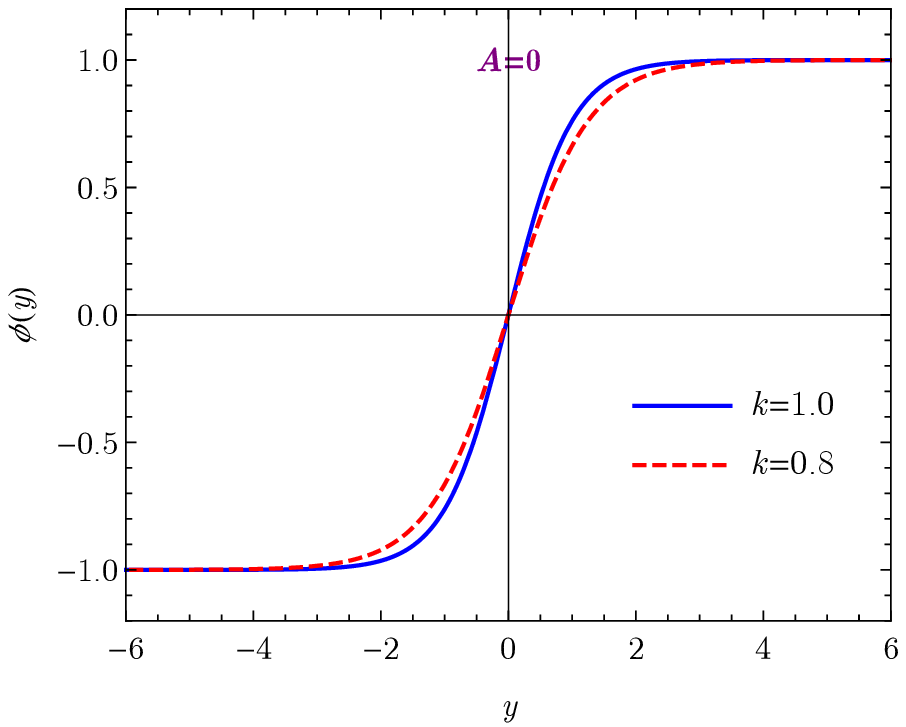}}
\subfigure[$~V(\phi)$]  {\label{Fig_Solution_A0_Scalar_Potential}
\includegraphics[width=7cm,height=5cm]{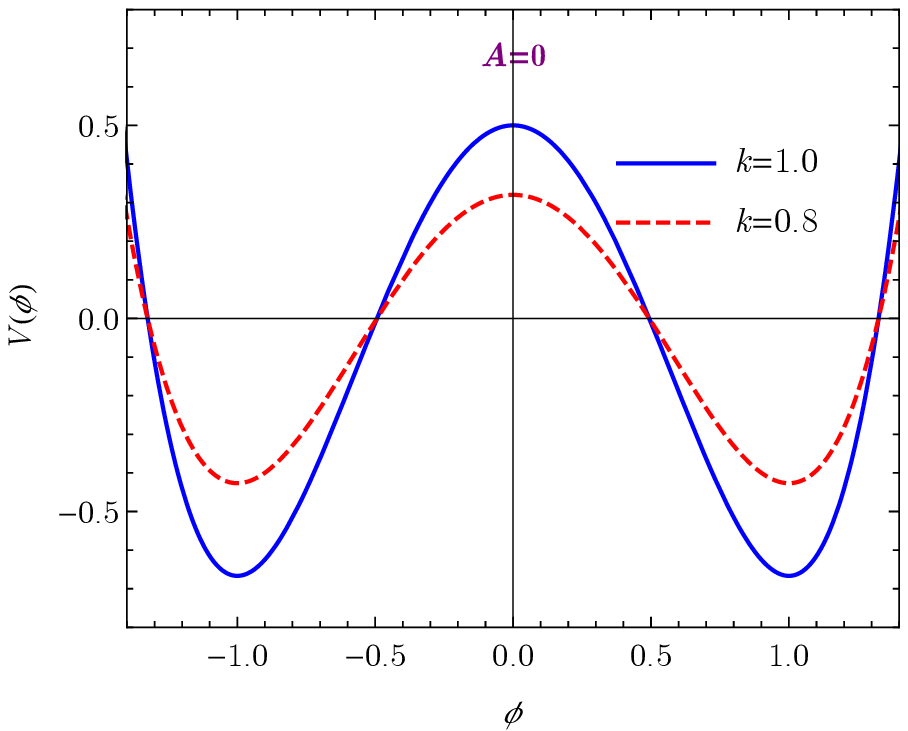}}
\subfigure[$~\rho(y)$]  {\label{Fig_Solution_A0_Energy_Density}
\includegraphics[width=7cm,height=5cm]{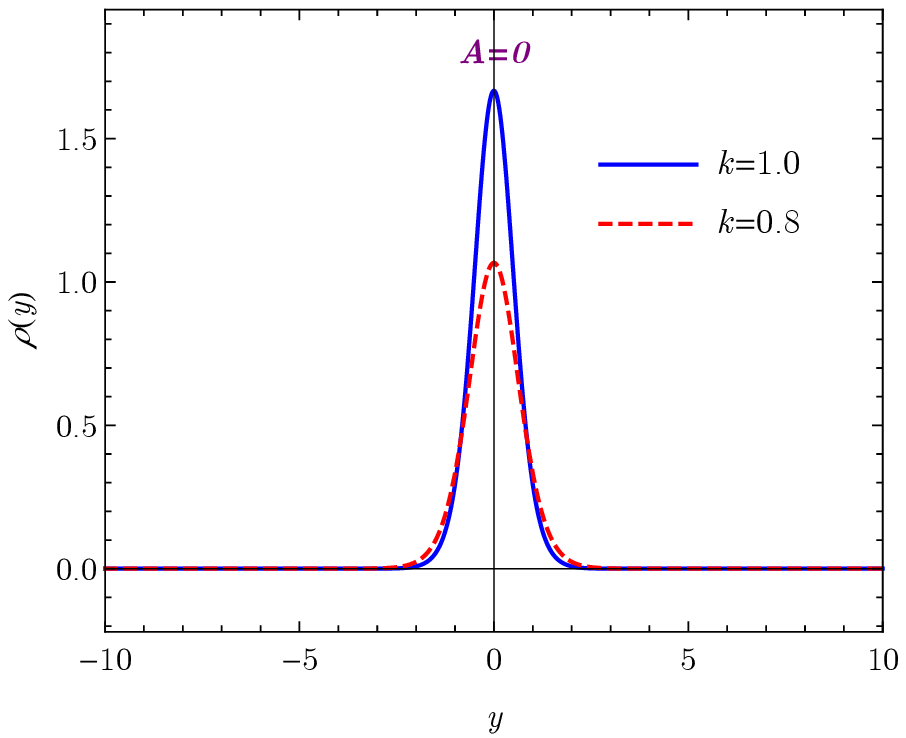}}
\end{center}
\caption{The profiles of the warp factor $a(y)$, scalar field $\phi(y)$, scalar potential $V(\phi)$, and energy density $\rho(y)$ for the case of  $A=0$.}
\label{Fig_Solution_A0}
\end{figure*}

 The second interesting case is $A=0$. Then from $A=12(\beta +2 \gamma )/\lambda$, $B=3 (2 \alpha +\beta +8\gamma)/\lambda$, and $\alpha+\beta+5\gamma=1$, one gets $B={6}/{\lambda }$, $\alpha=1-3 \gamma$, and $\beta= -2 \gamma$, with $\gamma$ a free parameter. Now the equations (\ref{EoM_Superpotential}) reduce to
\begin{subequations}\label{EoM_Superpotential_A0}
\beqn
a'&=&-\fc{aW}{3},\\
\phi'&=&\lt(\fc{1}{2}-\fc{W^2}{\lambda} \rt)W_\phi,\\
V&=&\lt(\fc{1}{4}-\fc{W^2}{2\lambda} \rt)  \phi 'W_\phi-\lt(\fc{1}{3}-\fc{W^2}{3\lambda} \rt)W^2.
\eeqn
\end{subequations}

For the superpotential ansatz $W(\phi)=2 k \phi$ with $\lambda\equiv 8 k^2$, a set of analytical solutions are obtained
 \begin{subequations}\label{Solution_A0}
 \beqn
a(y)&=&\text{sech}^{\frac{2}{3}}(k y),\\
 \phi (y)&=&\tanh (k y),\\
V( \phi)&=& \frac{k^2}{6}  \lt( 3-32 \phi ^2+64 \phi ^4 \rt),
 \eeqn
\end{subequations}
 where the integration constants have been chosen such that $a(0)=1$ and $\phi(0)=0$. The solutions of $a(y)$, $\phi(y)$, and $V(\phi)$ are illustrated  in Figs.~\ref{Fig_Solution_A0_Warp_Factor}, \ref{Fig_Solution_A0_Scalar_Field}, and \ref{Fig_Solution_A0_Scalar_Potential}, respectively. The warp factor is $Z_2$ symmetric, and the scalar field is a kink with $\phi (\pm \infty)\to\pm 1$, which connects two local minima of the scalar potential, i.e., $V_0=V(\pm 1)=-{k^2}/{96}$.

In this case, the torsion scalar reads $T=-\frac{16k^2}{3} \tanh ^2(k y)$, and it approaches $-\fc{16 k^2}{3}$ at the vacua $y\to\pm\infty$.  Therefore, the spacetime torsion is also enhanced as the parameter $k$ goes larger. As shown in Fig.~\ref{Fig_Solution_A0_Energy_Density}, the energy density of the brane is localized at $y=0$, and its thickness becomes thinner as the spacetime torsion gets stronger as well.

 \subsection{Case $B=0$}

\begin{figure*}[htb]
\begin{center}
\subfigure[$~a(y)$]  {\label{Fig_Solution_B0_Warp_Factor}
\includegraphics[width=7cm,height=5cm]{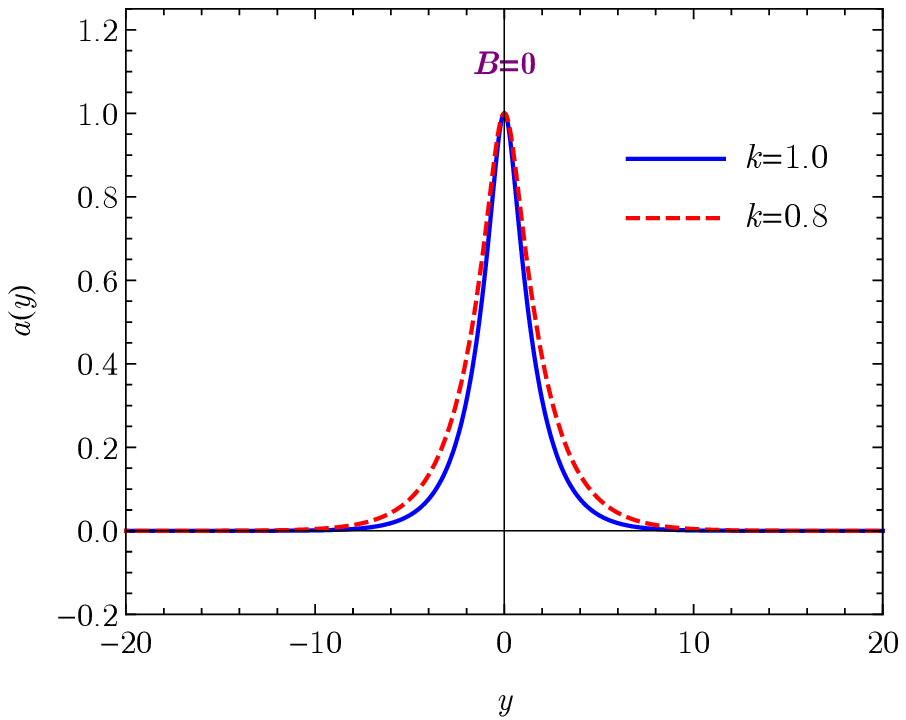}}
\subfigure[$~\phi(y)$]  {\label{Fig_Solution_B0_Scalar_Field}
\includegraphics[width=7cm,height=5cm]{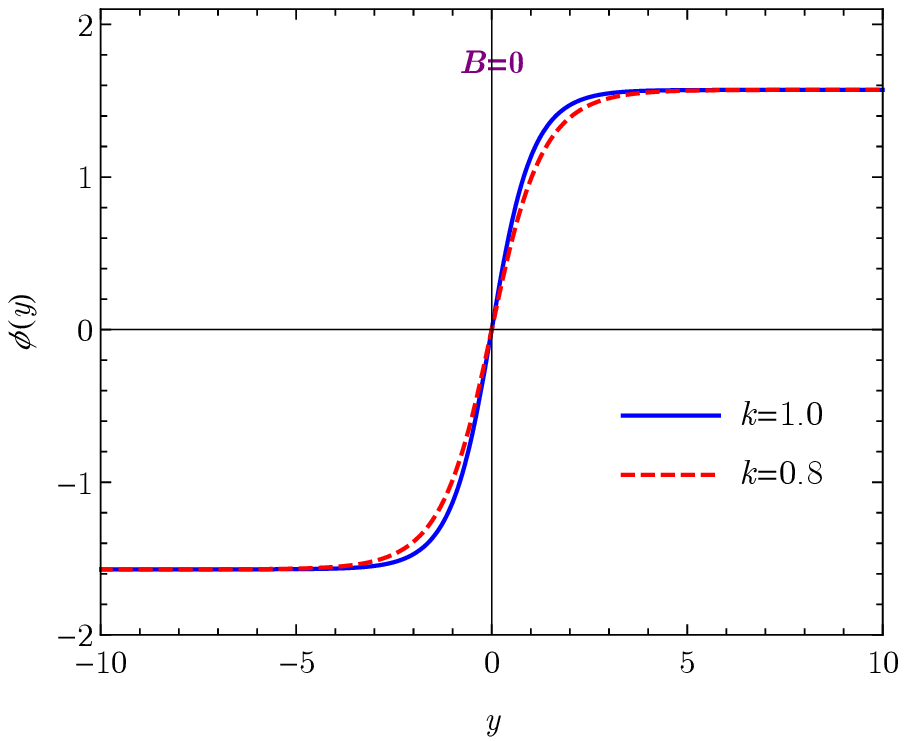}}
\subfigure[$~V(\phi)$]  {\label{Fig_Solution_B0_Scalar_Potential}
\includegraphics[width=7cm,height=5cm]{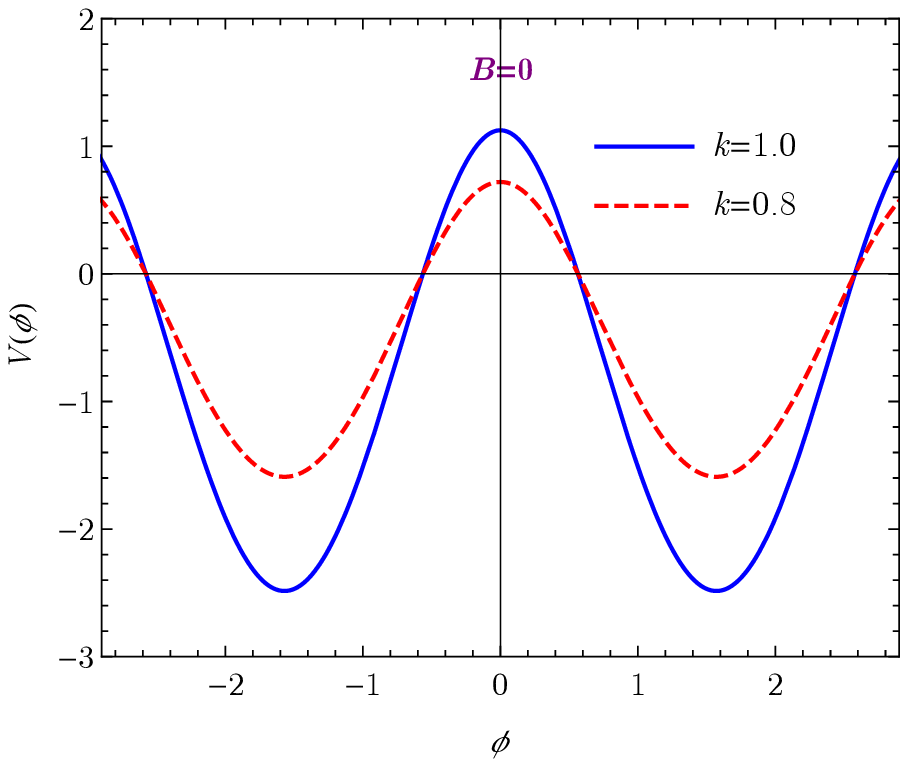}}
\subfigure[$~\rho(y)$]  {\label{Fig_Solution_B0_Energy_Density}
\includegraphics[width=7cm,height=5cm]{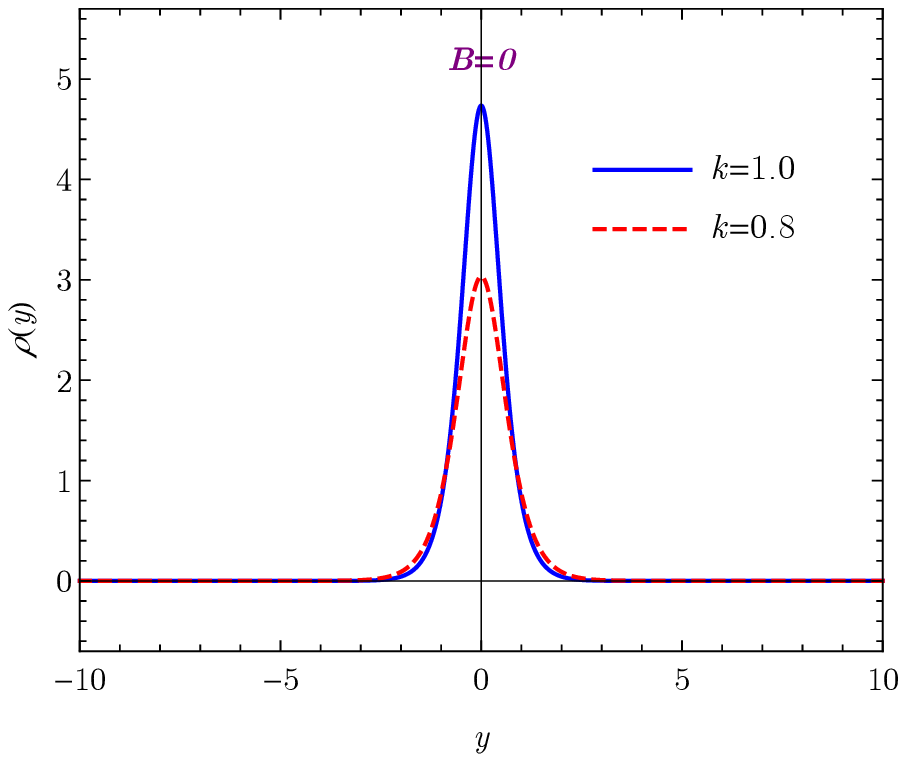}}
\end{center}
\caption{The profiles of the warp factor $a(y)$, scalar field $\phi(y)$, scalar potential $V(\phi)$, and energy density $\rho(y)$ for the case of $B=0$.}
\label{Fig_Solution_B0}
\end{figure*}

The last interesting case is $B=0$. Then from $A=12(\beta +2 \gamma )/\lambda$, $B=3 (2 \alpha +\beta +8\gamma)/\lambda$, and $\alpha+\beta+5\gamma=1$, one has $A={24}/{\lambda }$, $\alpha=-(1+3 \gamma)$, and $\beta= 2(1-\gamma)$, with $\gamma$ a free parameter. In this case, the equations (\ref{EoM_Superpotential}) become
\begin{subequations}\label{EoM_Superpotential_B0}
\beqn
a'&=&-\fc{aW}{3},\\
\phi'&=&\frac{W_\phi}{2}\left(1-\frac{8 W^2}{3 \lambda }\right)^{-\fc{3}{2}},\\
V&=&\frac{1}{4}\Bigg[\phi ' W_\phi-\left(\lambda-\frac{8 W^2}{3}\right) \lt(1-\sqrt{1-\frac{8 W^2}{3 \lambda }}\rt) \Bigg]
\nn\\
&\times& \left(1-\frac{8 W^2}{3 \lambda }\right)^{-\fc{3}{2}}.
\eeqn
\end{subequations}
By assuming the superpotential as the form $W(\phi)=\frac{3 k \sin\phi}{\sqrt{1+\sin ^2\phi}}$, a set of analytical solutions can be achieved as
 \begin{subequations}\label{Solution_B0}
 \beqn
a(y)\!&=&\!\fc{\lt(1+\sqrt{2}\rt)^{\frac{\sqrt{2}}{3}}}
{ \lt({\cosh^\fc{1}{2}(3 k y)+\sqrt{2} \cosh \lt(3k y/2\rt)}\rt)^{\frac{\sqrt{2}}{3}} },\\
\phi (y)\!&=&\!\arcsin\lt[\tanh \lt(3k y/2\rt)\rt],\\
V( \phi)\!&=&\!\frac{3k^2}{16}  \lt[35+3 \cos (2 \phi )-32 \lt(1+\sin^2\phi \rt)^\fc{1}{2}\rt],\qquad
 \eeqn
\end{subequations}
where the integration constants have been chosen such that $a(0)=1$ and $\phi(0)=0$ as well. As shown in Fig.~\ref{Fig_Solution_B0}, the warp factor $a(y)$, scalar field $\phi(y)$, scalar potential $V(\phi)$, and brane energy density $\rho(y)$ have similar shapes with the previous two cases. The scalar field exhibits a  kink profile, whose two ends $\phi (\pm \infty)\to\pm \pi/2$ connect two vacua of the scalar potential, i.e., $V_0=V(\pm \pi/2)=6\lt(1-\sqrt{2}\rt) k^2$.

Now the torsion scalar reads $T=-6k^2[1-\text{sech}(3 k y)]$, and $T\to-6 k^2$ as $y\to\pm\infty$. The bulk torsion also gets  stronger as the parameter $k$ goes larger. As shown in Fig.~\ref{Fig_Solution_B0_Energy_Density}, the energy density of the brane is localized at the origin, and its thickness becomes thinner as the torsion becomes stronger as the previous cases.

 \section{Gravitational perturbation}\label{Perturbation}

The full perturbed metric against the domain wall backgrounds takes the form
\beqn
ds^2&=&g_{MN}dx^Mdx^N \nn\\
&=&a^2(y)\big[\eta_{\mu\nu}+2h_{\mu\nu}+2\eta_{\mu\nu}\varphi
+2\pt_\mu\pt_\nu B\nn\\
&+&2\pt_{(\mu} C_{\nu)}\big]dx^\mu dx^\nu+a(y)(\pt_\mu F+G_\mu)dx^\mu dy \nn\\
&+&(1+2\psi)dy^2,
\label{Perturbed_Metric}
\eeqn
where $h_{\mu\nu}$ is  the transverse-traceless (TT) tensor mode, $G_\mu$ and $C_\nu$ are the transverse vector modes, and $\psi$, $\varphi$, $F$, and $B$ are the scalar modes. Due to the broken 5-dimensional local Lorentz invariance of current theory, the broken gauge freedom in tangent frame will release 10 extra degrees of freedom in the vielbein \cite{Li2011d}. Thus, one can generally write the perturbed vielbein as the form $e^A{}_M=\bar{e}^A{}_M+\ae^A{}_M$ \cite{Wu2012a,Izumi2013}. The degrees of freedom of the perturbed metric $g_{MN}$ are encoded in ${\bar e}^A{}_{M}$, satisfying the condition $g_{MN}=\eta_{AB}{\bar e}^A{}_{M}{\bar e}^A{}_{N}$. Then, all the 10 extra degrees of freedom are included in ${\ae^A}_{M}$,  which can be decomposed explicitly as ${\ae^5}_{5}=0$, ${\ae^a}_{5}=\delta^{a\mu}(\pt_\mu \beta+D_\mu)$, ${\ae^5}_{\mu}=0$, ${\ae^a}_{\mu}=\delta^{a\nu}\epsilon_{\mu\nu\lambda}(\pt^\lambda \sigma+V^\lambda)$, with $\beta$ a scalar, $D_\mu$ a transverse vector, $\sigma$ a pseudo-scalar, and $V^\lambda$ a transverse pseudo-vector. After taking the scalar-vector-tensor decomposition in the linear perturbed equations of motion, the TT tensor,  transverse (pseu-)vector, and (pseu-)scalar modes can be decoupled from each other, and they can be dealt with separately.

Here, we focus on studying the property of 4-dimensional KK gravitons in current models. This is done by considering the TT tensor perturbation against the domain wall backgrounds. So we close all the transverse (pseu-)vector and (pseu-)scalar modes in the perturbed metric (\ref{Perturbed_Metric}). Now, the non-vanishing components of the perturbed inverse metric are $g^{\mu\nu}=a^{-2}\lt(\eta ^{\mu\nu }-2 h^{\mu\nu }\rt)$ and $g^{55}=1$.

Correspondingly, the perturbed vielbein with respect to the perturbed metric reads
\beq
{e^A}_M=\left( {\begin{array}{*{20}{c}}
a(y)\left(\delta^a{}_\mu+\eta^{a\nu}{h_{\nu\mu}} \right)&  0\\
0  & 1
\end{array}} \right).
\eeq
Then, the inverse of the vielbein is given by
 \beq
{e_A}^M=\left( {\begin{array}{*{20}{c}}
a^{-1}(y)\left(\delta_a{}^\mu-\eta_{a\nu}{h^{\nu\mu}} \right)&  0\\
0  & 1
\end{array}} \right).
\eeq

With the perturbed vielbein and metric, the expressions for the non-vanishing components of the perturbed torsion tensor $T^{P }{}_{MN}$ read
\begin{subequations}
\beqn
{T^{\lambda }}_{5 \nu }&=&-{T^{\lambda }}_{\nu 5}=H {\delta^{\lambda }}_{\nu }+{h'}^{\lambda}{}_{\nu },\\
{T^{\lambda }}_{\mu\nu }&=&\partial _{[\mu }h^{\lambda }{}_{\nu ]}.
\eeqn
\end{subequations}
Then, the non-vanishing components of the perturbed contorsion tensor $K^{P}{}_{MN}$ are given by
\begin{subequations}
\beqn
{K^{\lambda }}_{5 \nu }&=&-H {\delta ^{\lambda }}_{ \nu}-{h'}^{\lambda }{}_{ \nu},\\
{K^5}_{\mu\nu}&=&a^2 \lt( H \eta _{\mu\nu }+2 H h_{\mu\nu }+{h'}_{\mu\nu}\rt),\\
{K^{\lambda }}_{\mu\nu }&=&\partial ^{\lambda }h_{\mu\nu }-\pt_\mu{h^{\lambda }}_\nu.
\eeqn
\end{subequations}
With the expressions of the perturbed torsion tensor $T^{P }{}_{MN}$ and contorsion tensor $K^{P}{}_{MN}$, the non-vanishing components of the perturbed superpotential torsion tensor ${S_P}^{MN}$ read
\begin{subequations}
\beqn
{S_{\lambda }}^{\mu 0}&=&-{S_{\lambda }}^{0 \mu }=\frac{1}{2} \lt[(d-1) H \delta ^{\mu }{}_{\lambda }-{h'}^{\mu }{}_{\lambda }\rt],\\
{S_{\lambda }}^{\mu\nu }&=&\frac{1}{2a^{2}}\pt ^{[\mu }h^{\nu ]}{}_{\lambda }.
\eeqn
\end{subequations}

By taking the torsion tensor $T^{P }{}_{MN}$ and superpotential torsion tensor ${S_P}^{MN}$ into consideration, the perturbed tensor $F_{PQ} =\alpha F^{(1)}_{PQ}+\beta F^{(2)}_{PQ}+\gamma F^{(3)}_{PQ}$ can be calculated by the non-vanishing components of $F^{(1)}_{PQ}$, $F^{(2)}_{PQ}$, and $F^{(3)}_{PQ}$, given by
\begin{subequations}
\beqn
F^{(1)}_{\mu \nu }&=&(1-d) a^2 H^2 \lt[\eta _{\mu \nu }+2 h_{\mu \nu }+\frac{d-2}{d-1}\fc{ h'_{\mu \nu }}{H}\rt],\qquad\\
F^{(2)}_{\mu \nu }&=&\fc{1-d}{2}  a^2 H^2 \lt[\eta _{\mu \nu }+2 h_{\mu \nu }+\fc{d-2}{d-1}\fc{h'_{\mu \nu }}{H}\rt],\\
 F^{(2)}_{55}&=&\frac{d (1-d)}{2}  H^2,\\
F^{(3)}_{\mu \nu } &=& d (1-d)a^2 H^2 \lt(\eta _{\mu \nu }+2 h_{\mu \nu }\rt),\\
F^{(3)}_{55}&=&T=d (1-d) H^2.
\eeqn
\end{subequations}

Further, the perturbations of $\frac{\partial F_{PQ}}{ \partial {e^A}_M}$ and $\frac{\partial F_{PQ}}{\partial(\partial_S {e^A}_M)}$ can be assembled by  $\frac{\partial F_{PQ}}{ \partial {e^A}_M}=\alpha \frac{\partial F^{(1)}_{PQ}}{ \partial {e^A}_M}
+\beta \frac{\partial F^{(2)}_{PQ}}{ \partial {e^A}_M}+\gamma \frac{\partial F^{(3)}_{PQ}}{ \partial {e^A}_M}$ and $\frac{\partial F_{PQ}}{\partial(\partial_S {e^A}_M)}=\alpha \frac{\partial F^{(1)}_{PQ}}{\partial(\partial_S {e^A}_M)}+\beta \frac{\partial F^{(2)}_{PQ}}{\partial(\partial_S {e^A}_M)}+ \gamma \frac{\partial F^{(3)}_{PQ}}{\partial(\partial_S {e^A}_M)}$. Because there are too many items, the non-vanishing components of $\frac{\partial F^{(i)}_{PQ}}{ \partial {e^A}_M}$ and $\frac{\partial F^{(i)}_{PQ}}{\partial(\partial_S {e^A}_M)}$ are listed in \ref{appendix}.

With all of these perturbed terms, the linear perturbation of the field equations (\ref{EoM}) is obtained finally
\beq
F_2{h''}_{\mu\nu }+ F_1 {h'}_{\mu\nu } + F_0\Box^{(4)}h_{\mu\nu }=0,
\label{Perturbed_EoM}
\eeq
 where $\Box^{(4)}=\eta^{\rho\sigma}\pt_{\rho }\pt_{\sigma }$ is the 4-dimensional d'Alembert operator, and the functions $F_2$, $F_1$ and $F_0$ are defined as
\begin{subequations}\label{Fi_Functions}
 \beqn
 F_2\!&\!=\!&\! \fc{A}{d \lt(1-A H^2\rt)}+\fc{B}{1-B H^2}-\fc{(d-1) C^2 H^2}{2 \lt(1-B H^2\rt)^2},\\
F_1\!&\!=\!&\!\fc{A H}{1-A H^2}-\fc{ (d-1)\lt[2- (2 B-dC)H^2\rt]C H}{2 \lt(1-B H^2\rt)^2}\nn\\
\!&\!+\!&\!\fc{ \lt[d B+ (d-1)C\rt]H}{1-B H^2}-\fc{2 A B H H'}{\lt(1-A H^2\rt) \lt(1-B H^2\rt)}\nn\\
\!&\!+\!&\!\fc{A^2 H H'}{d \lt(1-A H^2\rt)^2}-\fc{\lt[(d-2)B^2+(d-1)C^2\rt] H H'}{\lt(1-B H^2\rt)^2}\nn\\
\!&\!+\!&\!\lt[\fc{A C^2}{2 \lt(1-A H^2\rt)}
+\fc{ (d-4)BC^2 }{2 \lt(1-B H^2\rt)}\rt]\fc{(d-1) H^3 H'}{ \lt(1-B H^2\rt)^2},\qquad\\
F_0\!&\!=\!&\!\frac{2 (d-1) }{\lambda  a^2}\lt(\fc{\gamma }{1-A H^2}+\fc{1-\gamma }{1-B H^2}\rt),
 \eeqn
\end{subequations}
 with $C=(d-2) (2 \alpha +\beta )/\lambda$.

In the low energy regime $\lambda\to\infty$, the leading orders of the coefficient functions $F_2$, $F_1$ and $F_0$ read $F_2\to {2 (d-1)}{\lambda}^{-1}$, $F_1\to {2 (d-1) d H}{\lambda}^{-1}$, and $F_0\to{2 (d-1)}{\lambda}^{-1}a^{-2}$, respectively. Then, the evolution equation (\ref{Perturbed_EoM}) reduces to the standard form in GR, i.e.,
\beq
{h''}_{\mu\nu }-4 H {h'}_{\mu\nu } + a^{-2}\Box^{(4)}h_{\mu\nu }=0.
\eeq

If $F_2/F_0>0$, we can make a coordinate transformation $dz=dy/\sqrt{F_2/F_0}$ to eliminate the prefactors $F_2$ and $F_0$ in Eq.~(\ref{Perturbed_EoM}). Then, the evolution equation (\ref{Perturbed_EoM}) is recast as
\beq
\ddot{h}_{\mu\nu }+F_3 \dot{h}_{\mu\nu } + \Box^{(4)}h_{\mu\nu }=0,
\label{Perturbed_EoM_2}
\eeq
where the dot denotes the derivative with respect to the coordinate $z$, and $F_3=\fc{F_1}{\sqrt{F_0 F_2}}+\fc{1}{2}\lt(\fc{\dot{F}_0}{F_0}- \fc{\dot{F}_2}{F_2}\rt)$.

In order to get a Schr\"odinger-like equation, we employ a KK decomposition as the from
$h_{\mu\nu}=\epsilon_{\mu\nu}(x^{\mu})\Phi (z)e^{-\int{\frac{F_3}{2} dz}}$. Then,  the evolution equation (\ref{Perturbed_EoM_2}) can be decompose into two equations. The first one is a Klein-Gordon equation $(\Box^{(4)}-m^2)\epsilon_{\mu\nu}=0$ owing to the preserved 4-dimensional Poincar\'e invariance, where $m$ is the observed 4-dimensional effective mass of KK gravitons. The other is the aimed Schr\"odinger-like equation
\beq
-\ddot\Phi (z) +U(z)\Phi (z) =m^2 \Phi (z),
\label{Schrodinger_Equation}
\eeq
where the effective potential $U(z)$ is given by
\beq
U(z)=\fc{\dot{F}_3}{2}+\fc{F_3^2}{4}.
\label{Effective_Potential}
\eeq
The Hamiltonian can be further factorized as a supersymmetric quantum mechanics form, i.e., $\mc{H}=\mc{A}^\dag \mc{A}=$\\
$\lt(\pt_z+\fc{F_3}{2}\rt)\lt(-\pt_z+\fc{F_3}{2}\rt)$. With the Neumann boundary condition $\pt_z h_{\mu\nu} =0$, the self-adjoint Hamiltonian leads to non-negative eigenvalues $m^2\geq 0$  \cite{Yang2017}. Therefore, the KK gravitons are tachyonic free. The wave function of the massless graviton can be directly derived from the equation $\lt(-\pt_z+\fc{F_3}{2}\rt)\Phi _0(z)=0$, yielding
\beq
\Phi _0(z)=N_0e^{\int{\frac{F_3}{2} dz}},
\eeq
where $N_0$ is the normalization constant.

\subsection{Case $A=B$}

\begin{figure*}[htb]
\begin{center}
\subfigure[$~U(z)$]  {\label{AB_Effective_Potential}
\includegraphics[width=7cm,height=5cm]{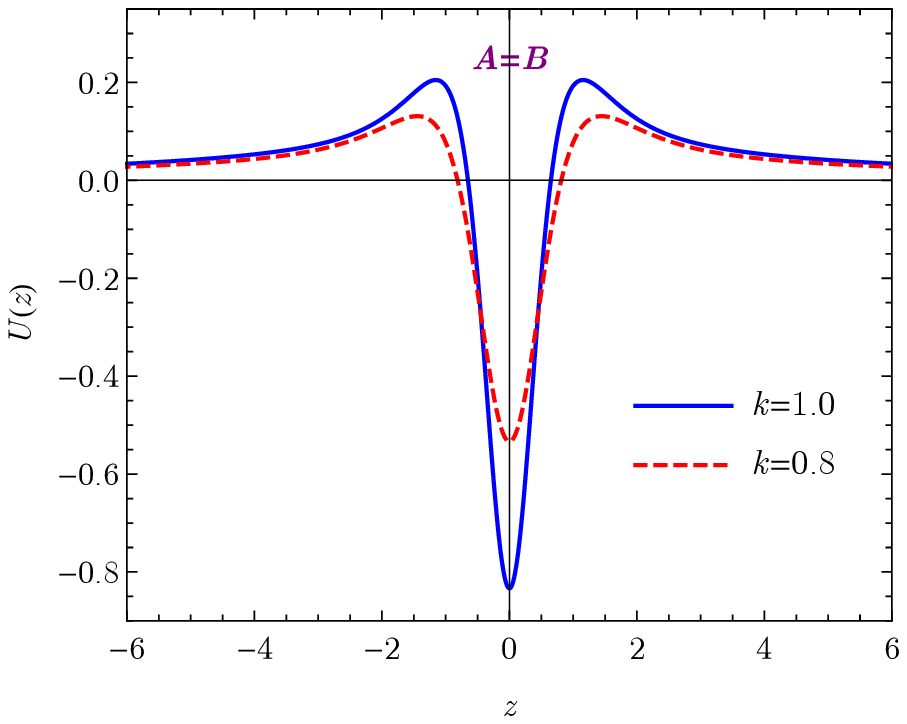}}
\subfigure[$~\Phi_0(z)$]  {\label{AB_Zero_Mode_Wave_Function}
\includegraphics[width=7cm,height=5cm]{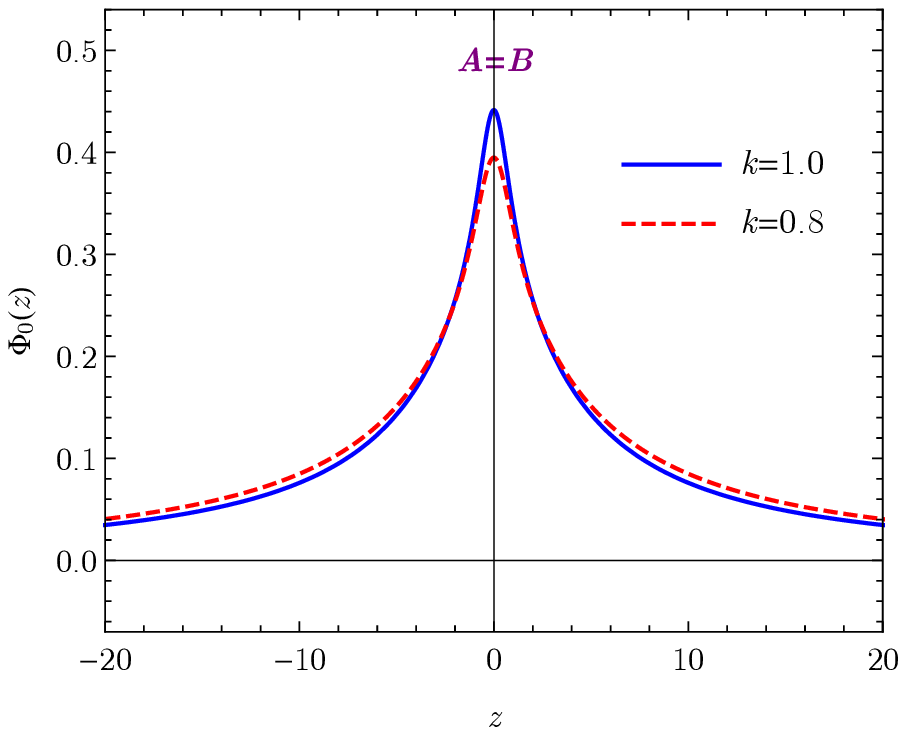}}
\end{center}
\caption{The profiles of the effective potential $U(z)$ and gravitational zero mode $\Phi_0(z)$ for the case of  $A=B$.}
\label{Fig_Perturbation_AB}
\end{figure*}

In this case, the parameters are $A=B=\frac{24}{5 \lambda}$ and $C=\fc{16-80 \gamma }{5 \lambda }$, with $\gamma$ a free parameter. It is easily verified that
\beq
\fc{F_2}{F_0}=\frac{a^2}{1-\phi^2}\lt[1-\lt(\frac{40 \gamma ^2}{3}-\frac{16 \gamma }{3}+\frac{23}{15}\rt) \phi^2\rt]\geq a^2.
\eeq
So the coordinate transformation $dz=dy/\sqrt{F_2/F_0}$ is robust. Following the above procedure, the KK gravitons are tachyonic free. This conclusion can also be seen directly by recasting  Eq.~(\ref{Perturbed_EoM}) into the form, ${h''}_{\mu\nu }+ \fc{F_1}{F_2} {h'}_{\mu\nu } +\mathcal{M}^2h_{\mu\nu }=0$, with $\mathcal{M}^2\equiv m^2{F_0}/{F_2}$ acting as an effective squared mass \cite{Fukushima2019}.

Due to the free parameter $\gamma$, the expressions can be simplified a lot by fixing $\gamma=1/5$, which corresponds to $C=0$. In this case, $F_2/F_0=a^2$. With the domain wall solutions (\ref{Solution_AB}), the functions $F_2$, $F_1$ and $F_0$ in Eqs.~(\ref{Fi_Functions}) are given explicitly as
\beqn
F_2\!\!&\!=\!&\!\!\frac{45}{16 k^2}\lt(3+\tanh ^2(k y)\rt),\\
F_1\!\!&\!=\!&\!\!-\frac{5 }{16 k}\lt[27 \text{sech}^2(k y)+8\sqrt{9+3\tanh ^2(k y)}\rt]\nn\\
&\times&\tanh (k y) ,\\
F_0\!\!&\!=\!&\!\!\frac{45 \left(3+\tanh ^2(k y)\right)}{16 k^2}\left[\left(2-\sqrt{3}\right) \cosh (k y)
\right.\nn\\
&&\times \left.\left(2+\sqrt{4-\text{sech}^2(k y)}\right)\right]^{\frac{2}{3 \sqrt{3}}}.
\eeqn

Since the coordinate transformation $dz=dy/a(y)$ cannot be integrated out analytically in this case, we cannot obtain the analytical expression of the effective potential in $z$ coordinate. However, the effective potential in $y$ coordinate can be expressed explicitly as
\beqn
U(y)\!&\!=\!&\!\frac{3 a^2}{4 \left(4 k^2-27 H^2\right)^2}\Big[5 H^2 \lt(16 k^4+729 H^4\rt)\nn\\
&-&270 H^4 \left(4 k^2-27 H'\right)+8 k^2  \lt(4 k^2-27 H'\rt)H'\nn\\
\!&\!-\!&\!81 H^2 H' \lt(16 k^2-9 H'\rt)-54  \lt(4 k^2-27 H^2\rt)H H''\Big]\nn\\
&=&-\frac{k^2 \text{sech}^8(k y) }{72\sqrt{4-\text{sech}^2(k y)} \left(3+\tanh ^2(k y)\right)^2 }\nn\\
&\times& \fc{ \cosh (k y)^{-\frac{2}{3 \sqrt{3}}} \left(2+\sqrt{4-\text{sech}^2(k y)}\right)^{-\frac{2}{3 \sqrt{3}}}}{ \left(2-\sqrt{3}\right)^{\frac{2}{3 \sqrt{3}}}\left(2+\sqrt{4-\text{sech}^2(k y)}\right)^2}\nn\\
&\times&\Bigg[4 \lt(1+2 \cosh (2 k y)\rt) \Big[5 \lt(28+15 \sqrt{3}\rt) \cosh (2 k y)\nn\\
&-& \left(113+6 \sqrt{3}\right) \cosh (4 k y)-5 \cosh (6 k y)+302\nn\\
&+&57 \sqrt{3}\Big]
+\sqrt{4-\text{sech}^2(k y)}\Big[2\lt(593+150 \sqrt{3}\rt)\nn\\
&+&2 \left(701+222 \sqrt{3}\right) \cosh (2 k y)-10 \cosh (8 k y)\nn\\
&-&3 \left(23-44 \sqrt{3}\right) \cosh (4 k y)-\left(12 \sqrt{3}+241\right) \nn\\
&\times&\cosh (6 k y)\Big] \Bigg].
\eeqn

Correspondingly, the wave function of the massless graviton in $y$ coordinate is given by
\beqn
\!\!\!\!\Phi _0(y)&=&N_0 a^{\fc{3}{2}}\left(1-\frac{27 H^2}{4 k^2}\right)^{3/4}\nn\\
&=&\frac{ N_0 3^{3/4}  \cosh ^{\frac{3}{2}}(k y) }{\left(2-\sqrt{3}\right)^{\frac{1}{2\sqrt{3}}}\lt(1+2 \cosh (2 k y)\rt)^{3/4}}
\nn\\
&\times& \left(2 \cosh (k y)+\sqrt{1+2 \cosh (2 k y)}\right)^{-\frac{1}{2 \sqrt{3}}}.
\eeqn

By numerically integrating out the coordinate transformation, the effective potential $U(z)$ and the wave function $\Phi_0(z)$ of the massless graviton in $z$ coordinate are illustrated in Fig.~\ref{Fig_Perturbation_AB}. As shown in Fig.~\ref{AB_Zero_Mode_Wave_Function}, the massless graviton is normalizable and its width decreases with the enhancement of the spacetime torsion. With the normalization condition $\int^{\infty}_{-\infty}{\Phi_0^2dz}=\int^{\infty}_{-\infty}{\fc{\Phi_0^2dy}{\sqrt{F_2/F_0}}}=1$, the normalization constant can be approximately calculated, yielding $N_0 \approx \sqrt{\fc{k}{4.681}}$. The normalizable massless mode ensures that the effective 4-dimensional gravity can be recovered on the brane. Further, by counting the contribution of the massless mode sector in the action (\ref{Full_Action}), the 4-dimensional gravitational constant can be derived from the reduction $\fc{1}{G_4}=\fc{1}{G_5N_0^2}\int^{\infty}_{-\infty}{\Phi_0^2dz}$. So the fundamental 5-dimensional gravitational constant $G_5$ reads $G_5={G_4}/N_0^2 \approx 4.681{G_4}/{k}$.

As shown in Fig.~\ref{AB_Effective_Potential}, the effective potential is volcano-like and the width of potential well decreases with the enhancement of the spacetime torsion. Therefore, the massless graviton is the only bound state, and the massive excited states possess a continuous mass spectrum. The massive states will contribute a correction to the Newtonian potential at short distance. Quantitatively, for two point-like sources separated by a distance $r$ on the brane, a volcano-like potential with $U(z)\sim\alpha(\alpha+1)/z^2$ for $z\gg 1$ will yield a correction proportional to ${1}/{(k r)^{2\alpha}}$ to the Newtonian potential  \cite{Csaki2000,Bazeia2009}. For the current case, the effective potential behaves like $U(z)\sim \frac{15}{4 z^2}$ for $z\gg 1$, so it leads to a correction proportional to ${1}/{(k r)^{3}}$ to the Newtonian potential.

 \subsection{Case $A=0$}

For case $A=0$, the parameters are $B=\frac{6}{\lambda}$ and $C=\fc{4-16 \gamma}{\lambda }$, with $\gamma$ a free parameter. It is easily shown that
\beq
\fc{F_2}{F_0}=\frac{9-\lt(32 \gamma ^2-16 \gamma +5\rt) \phi^2}{\lt(3- \phi^2\rt) \lt(3- \gamma\phi^2 \rt)}a^2.
\eeq
Due to $-1\leq\phi(y)\leq 1$, the ratio $F_2/F_0$ is positive everywhere in the intervals $\frac{1-\sqrt{3}}{4} <\gamma<\frac{1+\sqrt{3}}{4}$ or $\gamma>3$. So the coordinate transformation $dz=dy/\sqrt{F_2/F_0}$ is robust and the KK gravitons are tachyonic free in these parameter intervals.

The corresponding expressions can be simplified a lot by setting $C=0$, which is realized by fixing the free parameter $\gamma=1/4$. In this case, $F_2/F_0=\frac{12a^2}{12-\phi^2}$. With the solutions (\ref{Solution_A0}), the functions $F_2$, $F_1$ and $F_0$ in Eqs.~(\ref{Fi_Functions}) are written explicitly as
\beqn
F_2&=&\frac{9}{4 k^2 \left(3-\tanh ^2(k y)\right)},\\
F_1&=&-\frac{3 (11+4 \cosh (2 k y))\sinh (2 k y) }{4 k (2+\cosh (2 k y))^2},\\
F_0&=&\frac{3 (13+11 \cosh (2 k y))}{32 k^2 (2+\cosh (2 k y)) \text{sech}^{\frac{4}{3}}(k y)}.
\eeqn
Then, the effective potential in $y$ coordinate is given by
\beqn
U(y)\!&\!=\!&\!12 a^2 k^2\bigg[\frac{5 H^2}{16 k^2-3 H^2}
\nn\\
\!&\!-\!&\!\frac{9 \left(512 k^6-240 k^4 H^2 +24 k^2 H^4 +9 H^6\right) H'^2}{\left(4 k^2-3 H^2\right)^2 \left(16 k^2-3 H^2\right)^3}\nn\\
\!&\!+\!&\!\frac{8 \left(16 k^4-48 k^2 H^2+9 H^4\right) H'}{ \left(4 k^2-3 H^2\right)\left(16 k^2-3 H^2\right)^2}
\nn\\
\!&\!-\!&\!\frac{18 \left(4 k^2-H^2\right)H H''}{\left(4 k^2-3 H^2\right) \left(16 k^2-3 H^2\right)^2}\bigg]\nn\\
\!&\!=\!&\!-\frac{k^2 \text{sech}^{\frac{4}{3}}(k y)}{4 (2+\cosh (2 k y))^2 (13+11 \cosh (2 k y))^3}
\nn\\
&\times \!&\!
\Big[295290+365898 \cosh (2 k y)+56976 \cosh (4 k y)
\nn\\
\!&\!-\!&\!24669 \cosh (6 k y)-8602 \cosh (8 k y)
\nn\\
\!&\!-\!&\!605 \cosh (10 k y)  \Big].
\eeqn
\begin{figure*}[htb]
\begin{center}
\subfigure[$~U(z)$]  {\label{A0_Effective_Potential}
\includegraphics[width=7cm,height=5cm]{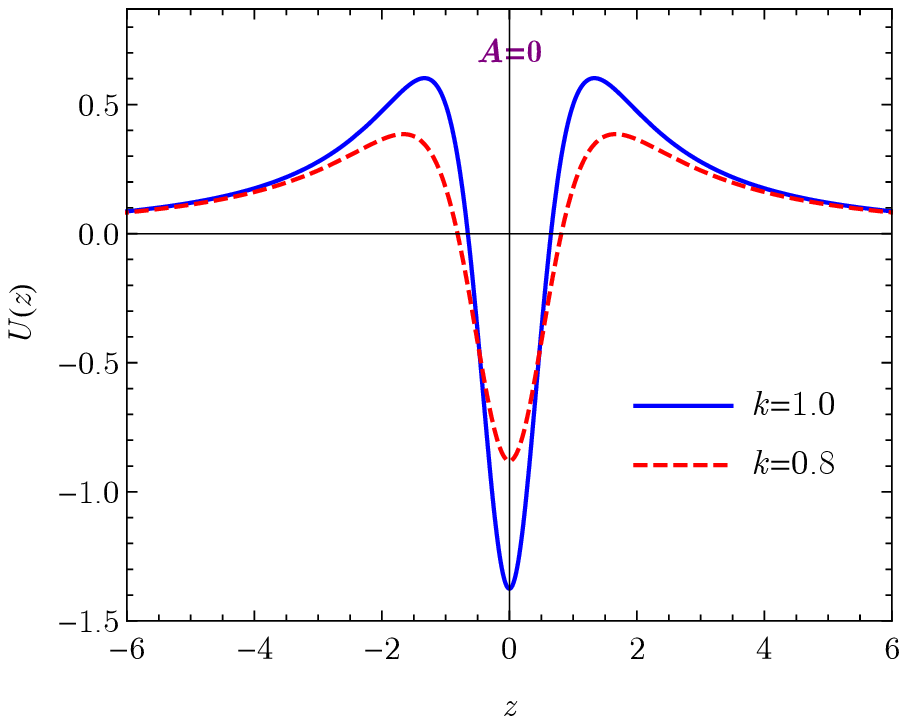}}
\subfigure[$~\Phi_0(z)$]  {\label{A0_Zero_Mode_Wave_Function}
\includegraphics[width=7cm,height=5cm]{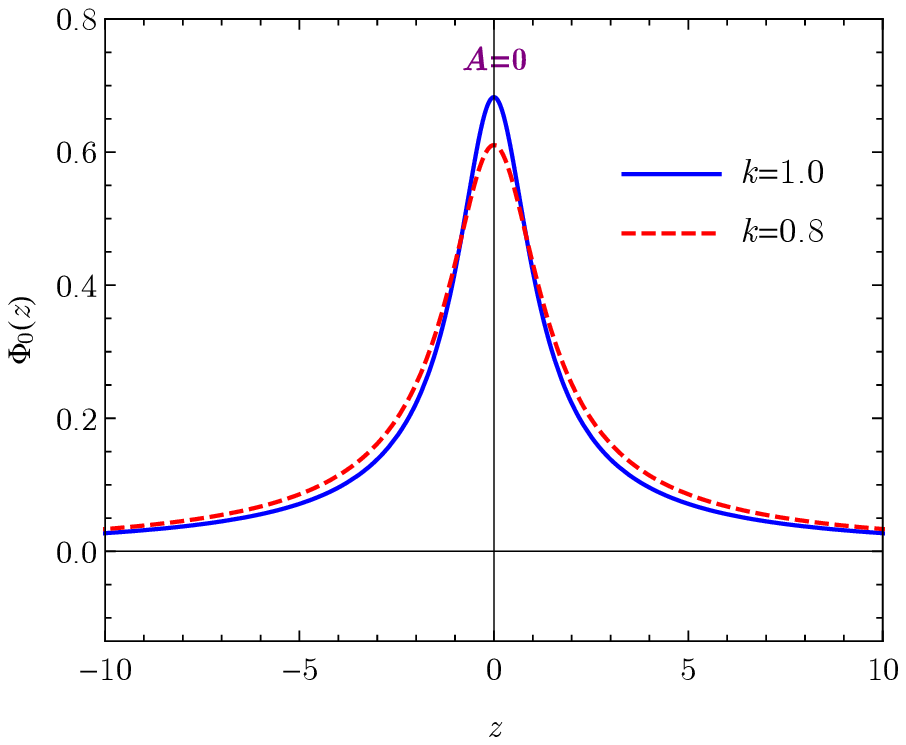}}
\end{center}
\caption{The profiles of the effective potential $U(z)$ and gravitational zero mode $\Phi_0(z)$ for the case of  $A=0$.}
\label{Fig_Perturbation_A0}
\end{figure*}
Moreover, the wave function of the massless graviton in $y$ coordinate is obtained as
\beqn
\Phi _0(y)&=&N_0 a^{\fc{3}{2}} \lt(1-\frac{3 H^2}{4 k^2}\rt)^{\fc{1}{2}} \lt(1-\frac{3 H^2}{16 k^2}\rt)^{\fc{1}{4}}\nn\\
&=&N_0 \text{sech}(k y)\lt( 1-\frac{\tanh ^2(k y)}{3} \rt)^{\fc{1}{2}}
\nn\\
&\times &\lt(1-\frac{\tanh ^2(k y)}{12} \rt)^{\fc{1}{4}},
\eeqn
where the normalization constant reads $N_0 =\sqrt{\frac{91 k}{453 \sqrt{\pi } }\fc{ \Gamma \left({1}/{6}\right)}{\Gamma \left({2}/{3}\right)}}\approx \sqrt{\fc{k}{2.146}}$. Therefore, the fundamental 5-dimensional gravitational constant is ${G_5}=G_4/N_0^2 \approx 2.146{G_4}/{k}$ in this model. The effective potential $U(z)$ and the wave function of the massless graviton $\Phi_0(z)$ in $z$ coordinate are illustrated in Fig.~\ref{Fig_Perturbation_A0}. It is shown that the widths of the potential well and wave function are both decreasing with the enhancement of the spacetime torsion. The volcano-like effective potential also behaves like $U(z)\sim \frac{15}{4 z^2}$ for $z\gg 1$, so it leads to a correction proportional to ${1}/{(k r)^{3}}$ to the Newtonian potential as the previous case.

\subsection{Case $B=0$}

In this case, the parameters are $A=\frac{24}{\lambda}$ and $C=-\fc{16 \gamma}{\lambda }$, with $\gamma$ a free parameter. Then, the ratio $F_2/F_0$ reads
\beq
\fc{F_2}{F_0}=\fc{3+\lt(6-8 \gamma ^2\rt) \sin^2\phi+3 \sin^4\phi }{3 \lt(1+\sin ^2\phi\rt) \lt(1+ \gamma  \sin^2\phi \rt)}a^2.
\label{Ratio_B0}
\eeq
Due to $ -\fc{\pi}{2}\leq \phi (y)\leq  \fc{\pi}{2}$, one has $0\leq \sin^2\phi(y)\leq 1$. Thus, the ratio $F_2/F_0$ is positive everywhere only if $\gamma<-\sqrt{\frac{3}{2}}$ or $-1<\gamma<\sqrt{\frac{3}{2}}$. The coordinate transformation $dz=dy/\sqrt{F_2/F_0}$ is robust and the KK gravitons are tachyonic free in these two parameter intervals.

In order to simplify the expressions, it is convenient to set $\gamma=0$, which vanishes the parameter $C$. Now the ratio (\ref{Ratio_B0}) reduces to $F_2/F_0=a^2 \lt(1+\sin^2 \phi \rt)$. With the solutions (\ref{Solution_B0}), the functions $F_2$,  $F_1$ and $F_0$ are written explicitly as
\beqn
F_2&=&\fc{1+ \tanh ^2\lt(3 k y/2\rt)}{4 k^2},\\
F_1&=& -\frac{\tanh \lt(3 k y/2 \rt)}{8 k}
 \Big[8 \lt(1+\tanh ^2\lt(3 k y/2 \rt)\rt)^{\fc{1}{2}}\nn\\
 &-&3  \text{sech}^2\lt(3 k y/2\rt)\Big],\\
F_0&=&\frac{1}{4 k^2}\lt(\cosh^\fc{1}{2} (3 k y)+\sqrt{2} \cosh \lt(3 k y/2\rt)\rt)^{\frac{2 \sqrt{2}}{3}}
\nn\\
&\times &\left(\sqrt{2}-1\right)^{\frac{2 \sqrt{2}}{3}}.
\eeqn
\begin{figure*}[htb]
\begin{center}
\subfigure[$~U(z)$]  {\label{B0_Effective_Potential}
\includegraphics[width=7cm,height=5cm]{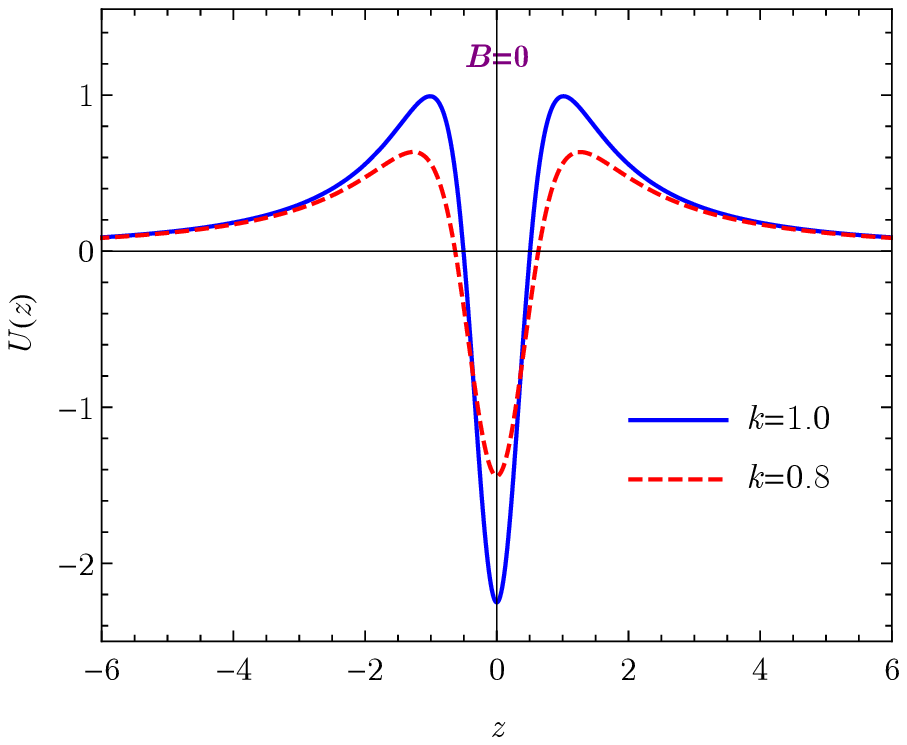}}
\subfigure[$~\Phi_0(z)$]  {\label{B0_Zero_Mode_Wave_Function}
\includegraphics[width=7cm,height=5cm]{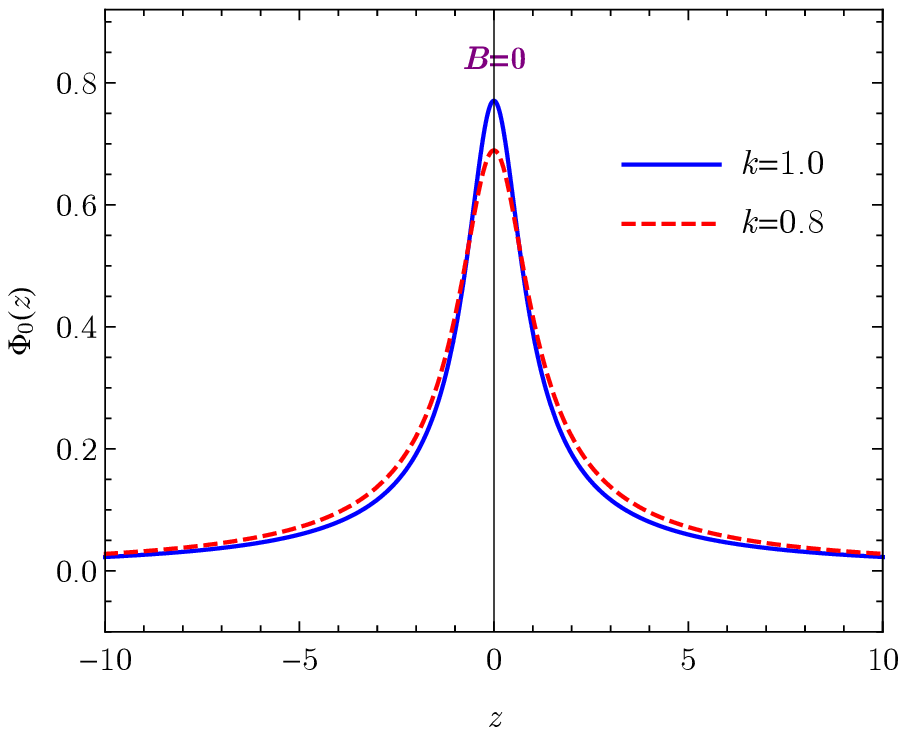}}
\end{center}
\caption{The profiles of the effective potential $U(z)$ and gravitational zero mode $\Phi_0(z)$ for the case of  $B=0$.}
\label{Fig_Perturbation_B0}
\end{figure*}
Then, the effective potential in $y$ coordinate can be expressed explicitly as
\beqn
\!\!U(y)\!\!&=&\!\!\frac{3 k^2 a^2 \lt(5 k^2 H^2 - 5 H^4 +2 k^2 H'\rt)}{4 \lt(k^2-H^2\rt)^2}\nn\\
&=&-\frac{5-5 \cosh (3 k y)+6 \lt(1+\tanh ^2\lt(3 k y/2\rt)\rt)}{\lt(\cosh^\fc{1}{2} (3 k y)+\sqrt{2} \cosh \lt(3 k y/2\rt)\rt)^{\frac{2 \sqrt{2}}{3}}}
\nn\\
&\times&\frac{3 k^2 \left(1+\sqrt{2}\right)^{\frac{2 \sqrt{2}}{3}} }{4 (1+\cosh (3 k y))} .
\eeqn
Correspondingly, the wave function of the massless graviton in $y$ coordinate is obtained as
\beqn
\Phi _0(y)\!&=&\!N_0 a^{\fc{3}{2}}\nn\\
\!&=&\! \fc{N_0\lt(1+\sqrt{2}\rt)^{\frac{1}{\sqrt{2}}} }{
\lt(\cosh^\fc{1}{2} (3 k y)+\sqrt{2} \cosh \lt(3 k y/2\rt)\rt)^{\frac{1}{\sqrt{2}}}},
\eeqn
where the normalization constant is estimated as $N_0 \approx \sqrt{\fc{k}{1.684}}$. Hence, the fundamental 5-dimensional gravitational constant reads ${G_5}=G_4/N_0^2 \approx 1.684{G_4}/{k}$. The effective potential $U(z)$ and the wave function of the massless graviton $\Phi_0(z)$ in $z$ coordinate are illustrated in Fig.~\ref{Fig_Perturbation_B0}. It is shown that the widths of the potential well and wave function are both decreasing with the enhancement of the spacetime torsion as the previous cases. The volcano-like effective potential behaves like $U(z)\sim \frac{15}{4 z^2}$ for $z\gg 1$ as well, which contributes a correction proportional to ${1}/{(k r)^{3}}$ to the Newtonian potential.

\section{Conclusions}\label{Conclusions}

We studied thick brane scenario in the context of the higher-dimensional BIDG in Weitzenb\"ock spacetime. In order to solve the equations of motion (\ref{EoM}) analytically, we resorted to the first-order formalism, which transforms the equations of motion into first-order equations by introducing a superpotential. Three analytic 3-brane solutions were obtained corresponding to some particular cases of parameter choices. It was found that all the three models describe the domain wall braneworld, where the kink scalar maps the boundaries of the extra dimension into two scalar vacua non-trivially. In every model, the thickness of the domain wall brane becomes thinner as the spacetime torsion gets stronger. Furthermore, by introducing a Yukawa coupling between the Dirac field and scalar field in the bulk, the massless chiral fermion can be localized on the domain wall brane \cite{Yang2018,Liu2017}.

Further,  we analyzed the tensor perturbation against the domain wall backgrounds and derived the Schr\"odinger-like equation (\ref{Schrodinger_Equation}) of KK modes. The effective potential of the Schr\"odinger-like equation is volcano-like, which is universal in flat brane scenario. The wave functions of the massless graviton are localized on the brane, so the effective 4-dimensional gravity can be recovered in each model. The widths of the potential well and wave function decrease with the enhancement of the spacetime torsion. The KK gravitons are tachyonic free for any $\gamma$ in the first model ($A=B$), for $\gamma\in\lt(\frac{1-\sqrt{3}}{4} ,\frac{1+\sqrt{3}}{4}\rt) \cup (3,\infty)$ in the second model ($A=0$), and for $\gamma\in \lt(-\infty,-\sqrt{\frac{3}{2}}\rt)\cup\lt(-1,\sqrt{\frac{3}{2}}\rt)$ in the third model ($B=0$).

Since the effective potentials of the Schr\"odinger-like equation have the same behavior like $U(z)\sim \frac{15}{4 z^2}$ for $z\gg 1$, the correction to the Newtonian potential is proportional to ${1}/{(k r)^{3}}$ in all modes. This is the same as the correction generated by RS2 model, but is different from the correction generated by domain wall branes in Eddington-inspired Born-Infeld gravity, where the correction is proportional to ${1}/{(k r)^{3+4n}}$ with $n$ a positive integer \cite{Liu2012,Fu2014}. The experimental test suggests that the length scale deviating from the gravitational inverse-square law is at least less than 48 $\mu$m \cite{Tan2020}, so the parameter $k$ is roughly estimated to be $k>10^{-3}$eV. From the relations ${G_5}\sim  {G_4}/{k}$ and $G_{d+1}=\lt(1/M_{d+1}\rt)^{d-1}$, a sparsity constraint on the fundamental 5-dimensional gravitational scale is obtained as ${M_5}>10^5$TeV.

Although the current scope of this works is to build the braneworld models and investigate the gravitational perturbation, the (pseudo-)scalar and (pseudo-)vector perturbations are also interesting and important. Due to the complexity of the field equations (\ref{EoM}), the calculation and analysis on these modes are left for our further investigation.

\section*{ACKNOWLEDGMENTS}

This work was supported by the National Natural Science Foundation of China under Grant No. 12005174. K. Yang acknowledges the support of Natural Science Foundation of Chongqing, China under Grant No. cstc2020jcyj-msxmX0370. H. Yu acknowledges the support of the Postdoctoral Science Foundation of Chongqing, China under Grant No. tc2021jcyj-bsh0124. Y. Zhong acknowledges the supported of the Natural Science Foundation of Hunan Province, China (Grant No. 2022JJ40033) and the Fundamental Research Funds for the Central Universities (Grants No.531118010195).

\appendix

\section{Perturbations of $\frac{\partial F_{PQ}}{ \partial {e^A}_M}$ and $\frac{\partial F_{PQ}}{\partial(\partial_S {e^A}_M)}$}
\label{appendix}

The perturbation of $\frac{\partial F_{PQ}}{ \partial {e^A}_M}$ can be assembled by  $\frac{\partial F_{PQ}}{ \partial {e^A}_M}=\alpha \frac{\partial F^{(1)}_{PQ}}{ \partial {e^A}_M}
+\beta \frac{\partial F^{(2)}_{PQ}}{ \partial {e^A}_M}+\gamma \frac{\partial F^{(3)}_{PQ}}{ \partial {e^A}_M}$, with the non-vanishing components of $\frac{\partial F^{(1)}_{PQ}}{ \partial {e^A}_M}$, $\frac{\partial F^{(2)}_{PQ}}{ \partial {e^A}_M}$, and $\frac{\partial F^{(3)}_{PQ}}{ \partial {e^A}_M}$ given by
\beqn
\frac{\pt {F}^{(1)}_{\mu\nu}}{\pt e^5{}_5}\!&\!=\!&\!2(d-1) a^2 H^2\Big[  \eta _{\mu\nu}+2  h_{\mu\nu}+\frac{d-2}{d-1} \frac{ {h'}_{\mu\nu}}{H}\Big],\\
\frac{\pt {F}^{(1)}_{5 \nu}}{\pt {e}^5{}_\lambda}\!&\!=\!&\!(1-d) H^2 \delta ^\lambda{}_\nu+(2-d) H {h'}^\lambda{}_\nu,\\
\frac{\pt {F}^{(1)}_{\mu 5}}{\pt{e}^a{}_5}\!&\!=\!&\!\frac{\pt {F}^{(1)}_{5 \mu }}{\pt{e}^a{}_5}=(1-d)a H^2 \Big[ \eta_{a \mu }+ h_{a\mu}+\frac{d-2}{d-1}\fc{ {h'}_{a\mu}}{H}\Big],\nn\\
&&\\
\frac{\pt {F}^{(1)}_{\mu\nu}}{\pt {e}^0{}_\lambda}\!&\!=\!&\!H\lt[{\pt_\nu h^\lambda{}_\mu}-(d-2) {\pt_\mu h^\lambda{}_\nu}+(d-3)\pt^\lambda h_{\mu\nu}\rt],\\
\frac{\pt {F}^{(1)}_{\mu\nu}}{\pt {e}^a{}_5}\!&\!=\!&\!a H \lt[\Big(\frac{3}{2}-d\Big) {\pt_\mu h_{a\nu}}+\frac{1}{2} {\pt_\nu h_{a\mu}}+(d-2) {\pt_a h_{\mu\nu}}\rt],\nn\\
&&\\
\frac{\pt{F}^{(1)}_{5\nu}}{\pt{e}^a{}_\lambda}\!&\!=\!&\!\frac{H}{a}\lt(\frac{1}{2}\pt ^\lambda h_{a \nu}+\frac{1}{2}{\pt_a h^\lambda{}_\nu}-{\pt_\nu h_a{}^\lambda}\rt),\\
\frac{\pt {F}^{(1)}_{\mu\nu}}{\pt{e}^a{}_\lambda}\!&\!=\!&\! a H^2 \lt[\Big(\frac{1}{2}-d\Big)  \delta^\lambda{}_\nu\eta _{a\mu}-\frac{1}{2}  \delta^\lambda{}_\mu\eta _{a\nu}+\delta _a{}^\lambda \eta _{\mu\nu}\rt]\nn\\
\!&\!+\!&\! a H^2 \Bigg[\Big(\frac{1}{2}-d\Big)  \delta ^\lambda{}_\nu h_{a\mu}-\frac{1}{2} \delta^\lambda{}_\mu h_{a\nu} - \delta _{\mu\nu} h_a{}^\lambda\nn\\
\!&\!+\!&\!  2 \delta _a{}^\lambda  h_{\mu\nu}\Bigg]- a H \Bigg[\frac{1}{2} \eta _{a\mu} {h'}^\lambda{}_\nu+\eta _{a\nu} {h'}^\lambda{}_\mu\nn\\
\!&\!-\!&\!\eta _{\mu\nu} {h'}_a{}^\lambda-\delta _a{}^\lambda {h'}_{\mu\nu}+\Big(d-\frac{3}{2}\Big) \delta^\lambda{}_\nu{h'}_{a\mu}  \Bigg],\\
\frac{\pt F^{(2)}_{\mu\nu}}{\pt {e}^5{}_5}\!&\!=\!&\!(d-1) a^2 H^2 \lt[ \eta _{\mu\nu}+2 h_{\mu\nu}+\frac{d-2}{d-1}\frac{ {h'}_{\mu\nu}}{H} \rt],\\
\frac{\pt F^{(2)}_{5 \nu}}{\pt {e}^5{}_\lambda}\!&\!=\!&\!\frac{1}{2} \lt[(1-d) H^2 \delta ^\lambda{}_\nu+(2-d) H {h'}^\lambda{}_\nu\rt],\\
\frac{\pt F^{(2)}_{\mu5}}{\pt {e}^5{}_\lambda}\!&\!=\!&\!(1-d) H^2 \delta ^\lambda{}_\mu+(2-d) H {h'}^\lambda{}_\mu,\\
\frac{\pt F^{(2)}_{\mu 5}}{\pt e^a{}_5}\!&\!=\!&\!\fc{\pt F^{(2)}_{5\mu}}{\pt e^a{}_5}=\fc{1-d}{2}a H^2  \Big[ \eta _{a\mu}+ h_{a\mu}+\frac{d-2}{d-1}\fc{ {h'}_{a\mu}}{H}\Big],\nn\\
&&\\
\frac{\pt F^{(2)}_{55}}{\pt e^a{}_\lambda}\!&\!=\!&\!(d-1) \fc{H^2 }{a}\lt[{\delta _a}^\lambda -  {h_a}^\lambda+\frac{d-2}{d-1 } \fc{{h'}_a{}^\lambda}{H} \rt],\\
\frac{\pt F^{(2)}_{\mu\nu}}{\pt e^5{}_\lambda}\!&\!=\!&\!\frac{H}{2}\bigg[{\pt_\mu h^\lambda{}_\nu}-(d-2) {\pt_\nu h^\lambda{}_\mu}+(d-3)\pt ^\lambda h_{\mu\nu}\bigg],\nn\\
&&\\
\frac{\pt F^{(2)}_{\mu\nu}}{\pt e^a{}_5}\!&\!=\!&\!\frac{a H}{2}  \bigg[\Big(\frac{3}{2}-d\Big) {\pt_\nu h_{a\mu}}+\frac{1}{2} {\pt_\mu h_{a\nu}} + (d-2) {\pt_a h_{\mu\nu}}\bigg],\nn\\
&&\\
\frac{\pt F^{(2)}_{\mu 5}}{\pt {e^a}_\lambda}\!&\!=\!&\!\frac{H}{2a}\lt[(d-3) {\pt_\mu {h_a}^\lambda}-(d-2) {\pt_a h^\lambda{}_\mu}+\pt^\lambda h_{a\mu}\rt],\nn\\
&&\\
\frac{\pt F^{(2)}_{5\nu}}{\pt e^a{}_\lambda}\!&\!=\!&\!\frac{H}{2a}\bigg[(d-2) {\pt_\nu h_a{}^\lambda}-\lt(d-\frac{3}{2}\rt) {\pt_a h^\lambda{}_\nu}+\frac{1}{2}\pt^\lambda h_{a\nu}\bigg],\nn\\
&&\\
\frac{\pt F^{(2)}_{\mu\nu}}{\pt {e}^a{}_\lambda}\!&\!=\!&\!\frac{a H^2 }{4} \Big[2  {\delta _a}^\lambda \eta _{\mu\nu} - (2 d-1) \delta_\mu{}^\lambda \eta _{a\nu} -  \delta _\nu{}^\lambda \eta _{a\mu} \Big]\nn\\
&-&\frac{a H^2}{4}  \Big[(2 d-1) {\delta _\mu}^\lambda h_{a \nu}+{\delta_\nu}^\lambda h_{a \mu} + 2 \eta _{\mu\nu} {h_a}^\lambda\nn\\
&-&4 {\delta _a}^\lambda h_{\mu\nu} \Big]-\fc{a H}{4}  \Big[\eta _{a \mu} {h'}^\lambda{}_\nu+2 (d-1){\delta_\mu}^\lambda {h'}_{a\nu}\nn\\
&+&{\delta_\nu}^\lambda {h'}_{a\mu}-2 \eta _{\mu\nu} {h'}_a{}^\lambda -2{\delta _a}^\lambda {h'}_{\mu\nu} \Big],\\
\fc{\partial F^{(3)}_{\mu\nu}}{\pt {e^5}_5}\!&\!=\!&\!2 d (d-1)a^2  H^2 \lt(\eta _{\mu\nu }+2 h_{\mu\nu }\rt),\\
\fc{\partial F^{(3)}_{\mu5}}{\partial {e^5}_\lambda}\!&\!=\!&\!\fc{ \pt F^{(3)}_{5\mu} }{\pt {e^5}_\lambda}=d (1-d) H^2 \delta ^{\lambda }{}_{\mu },\\
\fc{\partial F^{(3)}_{\mu 5 }}{{\pt e}^a{}_5}\!&\!=\!&\! \frac{\pt F^{(3)}_{5\mu }}{{\pt e}^a{}_5}= d (1-d) a H^2\lt(\eta _{a \mu }+h_{a \mu }\rt),\\
\frac{\pt F^{(3)}_{55}}{ {\pt e}^a{}_{\lambda}}\!&\!=\!&\! 2(d-1)\frac{H^2}{a} \lt[  \delta_{a}{}^{\lambda }- h_{a }{}^{\lambda }+\frac{d-2}{d-1}\frac{{h'}_{a }{}^{\lambda }}{H}\rt],\\
\frac{\pt F^{(3)}_{\mu\nu}}{\pt {e^a}_\lambda}\!&\!=\!&\! (1-d)a H^2 \lt(d \delta ^\lambda{}_{(\mu}\eta _{\nu)a}-2 \delta ^\lambda{}_a \eta _{\mu\nu}\rt)\nn\\
&+&(1-d)a H^2 \left(d \delta ^\lambda{}_{(\mu} h_{\nu)a}+2 \eta _{\mu\nu} h_a{}^\lambda-4 \delta _a{}^\lambda h_{\mu\nu} \rt)\nn\\
&+&2(d-2) a  H \eta _{\mu\nu} {h'}_a{}^\lambda.
\eeqn

The perturbation of $\frac{\partial F_{PQ}}{\partial(\partial_S {e^A}_M)}$ can be assembled by
$\frac{\partial F_{PQ}}{\partial(\partial_S {e^A}_M)}=\alpha \frac{\partial F^{(1)}_{PQ}}{\partial(\partial_S {e^A}_M)}+\beta \frac{\partial F^{(2)}_{PQ}}{\partial(\partial_S {e^A}_M)}+ \gamma \frac{\partial F^{(3)}_{PQ}}{\partial(\partial_S {e^A}_M)}$,where the non-vanishing components of $\frac{\partial F^{(1)}_{PQ}}{\partial(\partial_S {e^A}_M)}$, $\frac{\partial F^{(2)}_{PQ}}{\partial(\partial_S {e^A}_M)}$, and $\frac{\partial F^{(3)}_{PQ}}{\partial(\partial_S {e^A}_M)}$ are given by
\beqn
\frac{\pt {F}^{(1)}_{\mu5}}{\pt (\pt_5 e^5{}_\lambda)}\!&\!=\!&\! (1-d) H \delta ^\lambda{}_\mu+{h'}^\lambda{}_\mu,\\
\frac{\pt F^{(1)}_{\mu5}}{\pt ({\pt_\gamma e^5{}_5})}\!&\!=\!&\! (d-1) H \delta ^\gamma{}_\mu-{h'}^\gamma{}_\mu,\\
\frac{\pt {F}^{(1)}_{\mu\nu}}{\pt  ({\pt_5 e^5{}_\lambda)}}\!&\!=\!&\!\partial^\lambda h_{\mu\nu}-{\pt_\mu h^\lambda{}_\nu},\\
\frac{\pt F^{(1)}_{\mu\nu}}{\pt ({\pt_\gamma e^5{}_5})}\!&\!=\!&\!{\pt_\mu h^\gamma{}_\nu}-\pt^\gamma h_{\mu\nu},\\
\frac{\pt F^{(1)}_{5\nu}}{\pt ({\pt_5 e^a{}_\lambda})}\!&\!=\!&\!\frac{1}{2a}\lt({\pt_a h^\lambda{}_\nu}-\partial ^\lambda h_{a\nu}\rt),\\
\frac{\pt F^{(1)}_{5\nu}}{\pt (\pt_\gamma e^a{}_5)}\!&\!=\!&\!\frac{1}{2a}\lt(\partial ^\gamma h_{a\nu}-{\pt_a h^\gamma{}_\nu}\rt),\\
\frac{\pt F^{(1)}_{\mu5}}{\pt (\pt_\gamma e^5{}_\lambda)}\!&\!=\!&\!\fc{1}{a^{2}}\lt(\pt ^\gamma h^\lambda{}_\mu-\pt ^\lambda h^\gamma{}_\mu\rt),\\
\frac{\pt F^{(1)}_{5\nu}}{\pt (\pt_\gamma e^5{}_\lambda)}\!&\!=\!&\!\frac{1}{2a^{2}}\lt(\pt^\gamma h^\lambda{}_\nu-\pt ^\lambda h^\gamma{}_\nu\rt),\\
\frac{\pt F^{(1)}_{\mu\nu}}{\pt ({\pt_5 e^a{}_\lambda})}\!&\!=\!&\!-a \bigg[H\delta_a{}^\lambda\eta _{\mu\nu}+\lt(d-\frac{3}{2}\rt) H \delta ^\lambda{}_\mu\eta _{a\nu}\nn\\
&-&\frac{1}{2} H \delta ^\lambda{}_\nu \eta _{a\mu} \bigg]+a \bigg[H \eta _{\mu\nu} h_a{}^\lambda-2 H \delta _a{}^\lambda h_{\mu\nu}\nn\\
&+&\frac{1}{2}H \delta _\nu{}^\lambda h_{a\mu}-\lt(d-\frac{3}{2}\rt) H \delta ^\lambda{}_\mu h_{a\nu}+\eta _{a\nu } {h'}^\lambda{}_\mu\nn\\
&-&\delta _a{}^\lambda {h'}_{\mu\nu}+\frac{1}{2} \delta ^\lambda{}_\mu {h'}_{a\nu}+\frac{1}{2} \eta _{a\mu} {h'}^\lambda{}_\nu\bigg],\\
\frac{\pt F^{(1)}_{\mu\nu}}{\pt (\pt_\gamma e^a{}_5)}\!&\!=\!&\! -a H \bigg[\delta_a{}^\gamma \eta _{\mu\nu}-\frac{1}{2}  \delta ^\gamma{}_\nu \eta_{a\mu} +\lt(d-\frac{3}{2}\rt)  \delta ^\gamma{}_\mu\eta_{a\nu} \bigg]\nn\\
&+&a \bigg[H \eta _{\mu\nu} h_a{}^\gamma-2 H \delta _a{}^\gamma h_{\mu\nu}+\frac{1}{2}H \delta ^\gamma{}_\nu h_{a\mu}\nn\\
&-&\Big(d-\frac{3}{2}\Big) H \delta ^\gamma{}_\mu h_{a\nu}+\eta _{a\nu} {h'}^\gamma{}_\mu-\delta _a{}^\gamma {h'}_{\mu\nu}\nn\\
&+&\frac{1}{2} \delta _\mu{}^\gamma {h'}_{a\nu}+\frac{1}{2} \eta _{a\mu} {h'}^\gamma{}_\nu\bigg],\\
\frac{\pt F^{(1)}_{\mu\nu}}{\pt (\pt_\gamma e^5{}_\lambda)}\!&\!=\!&\!\frac{1}{2} \Big(H \delta^\gamma{}_\mu \delta^\lambda{}_\nu-H \delta ^\gamma{}_\nu \delta^\lambda{}_\mu+\delta^\gamma{}_\mu {h'}^\lambda{}_\nu\nn\\
&-&\delta ^\lambda{}_\mu {h'}^\gamma{}_\nu\Big),\\
\frac{\pt F^{(1)}_{5\nu}}{\pt (\pt_\gamma e^a{}_\lambda)}\!&\!=\!&\!\fc{1}{a}\Big(H \delta _a{}^\lambda \delta ^\gamma{}_\nu-H \delta _a{}^\gamma \delta ^\lambda{}_\nu+H \delta_\nu{}^\lambda h^\gamma{}_a\nn\\
&-&H \delta _\nu{}^\gamma  h^\lambda{}_a+\delta _a{}^\lambda {h'}^\gamma{}_\nu-\delta _a{}^\gamma {h'}^\lambda{}_\nu\Big),\\
\frac{\pt F^{(1)}_{\mu\nu}}{\pt (\pt_\gamma e^a{}_\lambda)}\!&\!=\!&\!\fc{1}{a}\bigg[\frac{1}{2}\delta^\lambda{}_\mu\lt(\pt ^\gamma h_{a\nu}-{\pt_a h^\gamma{}_\nu}\rt)\nn\\
&-&\frac{1}{2}\delta^\gamma{}_\mu\big(\pt ^\lambda h_{a\nu}-{\pt_a h^\lambda{}_\nu}\big)\nn\\
&+&\frac{1}{2}\eta _{a\mu}\lt(\pt ^\gamma h^\lambda{}_\nu-\pt ^\lambda h^\gamma{}_\nu\rt)\nn\\
&&+\eta _{a\nu}\big(\pt ^\gamma h^\lambda{}_\mu-\pt^\lambda h^\gamma{}_\mu\big)\nn\\
&+&\delta _a{}^\lambda\lt({\pt_\mu h^\gamma{}_\nu}-\pt ^\gamma h_{\mu\nu}\rt)\nn\\
&-&\delta _a{}^\gamma\lt({\pt_\mu h^\lambda{}_\nu}-\pt ^\lambda h_{\mu\nu}\rt)\bigg],\\
\frac{\pt F^{(2)}_{\mu5}}{\pt (\pt_5 e^5{}_\lambda)}\!&\!=\!&\!\frac{1}{2} \lt[(d-1) H \delta^\lambda {}_\mu-{h'}^\lambda{}_\mu\rt],\\
\frac{\pt F^{(2)}_{\mu5}}{\pt (\pt_\gamma e^5{}_5)}\!&\!=\!&\!\frac{1}{2} \lt[(1-d) H \delta^\gamma {}_\mu + {h'}^\gamma{}_\mu\rt],\\
\frac{\pt F^{(2)}_{55}}{\pt (\pt_5 e^a{}_\lambda)}\!&\!=\!&\!\frac{1-d}{a}\lt(H \delta _a{}^\lambda-H h_a{}^\lambda-\frac{{h'}_a{}^\lambda}{d-1}\rt),\\
\frac{\pt F^{(2)}_{55}}{\pt (\pt_\gamma e^a{}_5)}\!&\!=\!&\!\frac{d-1}{a}\lt( H \delta _a{}^\gamma-H h_a{}^\gamma-\frac{{h'}_a{}^\gamma}{d-1}\rt),\\
\frac{\pt F^{(2)}_{\mu\nu}}{\pt (\pt_5 e^5{}_\lambda)}\!&\!=\!&\!\frac{1}{2}\lt(\pt ^\lambda h_{\mu\nu}-{\pt_\nu h^\lambda{}_\mu}\rt),\\
\frac{\pt F^{(2)}_{\mu\nu}}{\pt (\pt_\gamma e^5{}_5)}\!&\!=\!&\!\frac{1}{2}\lt(\pt ^\gamma h_{\mu\nu}-{\pt_\nu h^\gamma {}_\mu}\rt),\\
\frac{\pt F^{(2)}_{\mu5}}{\pt ({\pt_5 e^a{}_\lambda})}\!&\!=\!&\!\frac{1}{2a}\lt({\pt_\mu h_a{}^\lambda}-\pt ^\lambda h_{a\mu}\rt),\\
\frac{\pt {F}^{(2)}_{5\nu}}{\pt  (\pt_5 e^a{}_\lambda)}\!&\!=\!&\!\frac{1}{4a}\left(2 {\pt_\nu h_a{}^\lambda}-\partial ^\lambda h_{a\nu}-{\pt_a h^\lambda{}_\nu}\right),\\
\frac{\pt F^{(2)}_{\mu 5}}{\pt (\pt_\gamma e^a{}_5)}\!&\!=\!&\!\frac{1}{2a}\lt(\partial^\gamma h_{a\mu}-{\pt_\mu h_a{}^\gamma}\rt),\\
\frac{\pt F^{(2)}_{5\nu}}{\pt (\pt_\gamma e^a{}_5)}\!&\!=\!&\!\frac{1}{4a} \lt(\pt^\gamma h_{a\nu}+{\pt_a h^\gamma{}_\nu}-2 {\pt_\nu h_a{}^\gamma}\rt),\\
\frac{\pt F^{(2)}_{5 \nu}}{\pt (\pt_\gamma e^5{}_\lambda)}\!&\!=\!&\!\frac{1}{4a^2}\lt(\partial ^\gamma h^\lambda{}_\nu-\partial ^\lambda h^\gamma{}_\nu \rt),\\
\frac{\pt F^{(2)}_{\mu\nu}}{\pt (\pt_\gamma e^5{}_\lambda)}\!&\!=\!&\!\frac{1}{4} \Big(H \delta _\mu{}^\lambda \delta_\nu{}^\gamma-H \delta _\mu{}^\gamma \delta_\nu{}^\lambda+\delta _\mu{}^\lambda {h'}^\gamma{}_\nu,\nn\\
&-&\delta _\mu{}^\gamma {h'}^\lambda{}_\nu\Big)\\
\frac{\pt {F}^{(2)}_{\mu\nu}}{\pt  (\pt_5 e^a{}_\lambda)}\!&\!=\!&\! -\frac{a}{2} \bigg[H \delta _a{}^\lambda \eta _{\mu\nu} +\Big(d-\frac{3}{2}\Big) H  \delta ^\lambda{}_\nu\eta _{a\mu}\nn\\
&-&\frac{1}{2}H  \delta ^\lambda{}_\mu \eta _{a\nu}\bigg]-\frac{a}{2}  \bigg[2 H \delta _a{}^\lambda h_{\mu\nu}-H \eta _{\mu\nu} h_a{}^\lambda\nn\\
&-&\frac{1}{2} H \delta^\lambda{}_\mu h_{a\nu}+\Big(d-\frac{3}{2}\Big) H \delta^\lambda{}_\nu h_{a\mu}+\delta _a{}^\lambda {h'}_{\mu\nu}\nn\\
&&-\delta ^\lambda{}_\nu {h'}_{a\mu}-\frac{1}{2} \delta^\lambda{}_\mu {h'}_{a\nu}-\frac{1}{2} \eta _{a\mu} {h'}^\lambda{}_\nu\bigg],\\
\frac{\pt F^{(2)}_{\mu\nu}}{\pt (\pt_\gamma e^a{}_5)}\!&\!=\!&\!\frac{a}{2}  \bigg[H \delta _a{}^\gamma \eta _{\mu\nu}+\lt(d-\frac{3}{2}\rt) H  \delta^\gamma{}_\nu\eta _{a\mu}\nn\\
&-&\frac{1}{2}H\delta^\gamma{}_\mu \eta _{a\nu}\bigg]+\frac{a}{2}  \bigg[2 H \delta _a{}^\gamma h_{\mu\nu}-H \eta _{\mu\nu} h_a{}^\gamma\nn\\
&-&\frac{1}{2} H\delta_\mu{}^\gamma h_{a\nu}+\lt(d-\frac{3}{2}\rt) H \delta_\nu{}^\gamma h_{a\mu}+\delta _a{}^\gamma {h'}_{\mu\nu}\nn\\
&-&\delta _\nu {}^\gamma {h'}_{a\mu} -\frac{1}{2} \delta_\mu{}^\gamma {h'}_{a\nu}-\frac{1}{2} \eta_{a\mu} {h'}^\gamma{}_\nu\bigg],\\
\frac{\pt F^{(2)}_{\mu\nu}}{\pt (\pt_\gamma e^a{}_\lambda)}\!&\!=\!&\!\frac{1}{4a}\Big[2\delta _a{}^\lambda \big({\pt_\nu h^{\gamma}{}_\mu}-\pt ^{\gamma}h_{\mu\nu}\big)\nn\\
&-&2\delta _a{}^{\gamma} \big({\pt_\nu h^\lambda{}_\mu}-\pt^\lambda h_{\mu\nu}\big)\nn\\
&+&\eta _{a\mu}\big(\pt ^{\gamma}h^\lambda{}_\nu-\pt ^\lambda h^{\gamma}{}_\nu\big)\nn\\
&&+\delta _\mu{}^{\gamma}\big(2 {\pt_\nu h_a{}^\lambda}-\pt ^\lambda h_{a\nu}-{\pt_a h^\lambda{}_\nu}\big)\nn\\
&-&\delta _\mu{}^\lambda\big(2 {\pt_\nu h_a{}^{\gamma}}-\pt ^{\gamma}h_{a\nu}-{\pt_a h^{\gamma}{}_\nu}\big)\nn\\
&+&2 \delta ^{\gamma }{}_\nu\big({\pt_\mu h_a{}^\lambda}-\pt ^\lambda h_{a\mu}\big)\nn\\
&-&2 \delta ^\lambda{}_\nu\big({\pt_\mu h_a{}^{\gamma}}-\pt ^{\gamma} h_{a\mu}\big)\Big],\\
\frac{\pt F^{(3)}_{55}}{\pt ({\pt_5 e^a{}_\lambda})}\!&\!=\!&\! \frac{2(1-d) }{a} \lt[H \delta _a{}^\lambda- H h_a{}^\lambda-\frac{{h'}_a{}^\lambda}{d-1}\rt],\\
\frac{\pt F^{(3)}_{55}}{\pt (\pt_\gamma e^a{}_5)}\!&\!=\!&\! \frac{2(d-1)}{a}\lt[H \delta _a{}^\gamma- H h_a{}^\gamma-\frac{{h'}_a{}^\gamma}{d-1}\rt],\\
\frac{\pt F^{(3)}_{55}}{\pt ({\pt_\gamma e^a{}_\lambda})}\!&\!=\!&\! \fc{2}{a^{3}}\lt(\pt^\gamma h_a{}^\lambda -\pt^{\lambda }h_a{}^\gamma \rt),\\
\frac{\pt F^{(3)}_{\mu\nu}}{\pt ({\pt_5 e^a{}_\lambda})}\!&\!=\!&\! 2 a \eta _{\mu\nu} \lt[(1-d) H \delta_a{}^\lambda-(1-d) H h_a{}^\lambda + {h'}_a{}^\lambda\rt]\nn\\
&+&4 (1-d) a H \delta _a{}^\lambda h_{\mu\nu},\\
\frac{\pt F^{(3)}_{\mu\nu}}{\pt ({\pt_\gamma e^a{}_5})}\!&\!=\!&\! 2 a \eta _{\mu\nu} \lt[(d-1) H \delta _a{}^\gamma-(d-1) H h_a{}^\gamma-{h'}_a{}^\gamma\rt]\nn\\
&+&4 (d-1) a H \delta _a{}^\gamma h_{\mu\nu},\\
\frac{\pt F^{(3)}_{\mu\nu}}{\pt ({\pt_\gamma e^a{}_\lambda})}\!&\!=\!&\!\fc{2}{a}\eta _{\mu\nu}\lt(\pt ^\gamma h_a{}^\lambda-\pt ^\lambda h_a{}^\gamma\rt).
\eeqn

\end{document}